\documentclass[preprint,authoryear,10pt]{elsarticle}

 \usepackage{graphics}
 \usepackage{hyperref}
 \usepackage{epstopdf}
 \usepackage{float}
\usepackage[table]{xcolor}

\usepackage{amssymb}
\usepackage{amsmath}
\usepackage{placeins}
\usepackage{natbib}

\usepackage[nodots]{numcompress}
\numberwithin{equation}{section}
\numberwithin{table}{section}
\numberwithin{figure}{section}
\nocite{*}

\biboptions{longnamesfirst,comma}

\begin{document}

\begin{frontmatter}



\title{Local semi-supervised approach to brain tissue classification in child brain MRI}


\author{Nataliya Portman\corref{cor1}}
\ead{nataliyaportman@gmail.com}

\author{Paule-J Toussaint}
\ead{paule.toussaint@mcgill.ca} 

\author{Alan C. Evans}
\ead{alan.evans@mcgill.ca}
\ead[url]{http://www.bic.mni.mcgill.ca/~alan}

\cortext[cor1]{Corresponding author}
\address{McConnell Brain Imaging Centre, Montreal Neurological Institute, McGill University, Montreal, QC, Canada}

\begin{abstract}
Most segmentation methods in child brain MRI are supervised and are based on global intensity distributions of major brain structures. The successful implementation of a supervised approach depends on availability of an age-appropriate probabilistic brain atlas. For the study of early normal brain development, the construction of such a brain atlas remains a significant challenge. Moreover, using global intensity statistics leads to inaccurate detection of major brain tissue classes due to substantial intensity variations of MR signal within the constituent parts of early developing brain.
\newline
In order to overcome these methodological limitations we develop a local, semi-supervised framework. It is based on Kernel Fisher Discriminant Analysis (KFDA) for pattern recognition, combined with an objective structural similarity index (SSIM) for perceptual image quality assessment. The proposed method performs optimal brain partitioning into subdomains having different average intensity values followed by SSIM-guided computation of separating surfaces between the constituent brain parts. The classified image subdomains are then stitched slice by slice via simulated annealing to form a global image of the classified brain.
\newline
In this paper, we consider classification into major tissue classes (white matter and grey matter) and the cerebro-spinal fluid and illustrate the proposed framework on examples of brain templates for ages 8 to 11 months and ages 44 to 60 months. We show that our method improves detection of the tissue classes by its comparison to state-of-the-art classification techniques known as Partial Volume Estimation. 
\end{abstract}


\end{frontmatter}


\section{Introduction}
\label{intro}
\subsection{Motivation}
The motivation for a new classification method comes from MRI study of early normal brain development \citep{Alm2007, Evans2006} designed to create a database of healthy controls comprised of T1-weigthed (T1w), T2-weighted (T2w) and proton-density weighted (pdw) images. 
MR dataset of early brain development (from birth to 4.5 years of age) poses a challenge to using automatic classification procedures such as FAST \citep{Zhang2001}, ANIMAl+INSECT \citep{ColZij1999} and SPM \citep{AshFr2005}. These methods rely on young adult brain atlases that do not adequately reflect dynamic changes in brain anatomy through early childhood   \citep{Alm2007, Fon2011}.
\newline
Furthermore, there is a number of factors inherent to early developing brain MR images that hamper automated classification process. They include high tissue intensity variation, low contrast-to-noise ratio between tissue types and partial volume artifact meaning that a brain voxel posesses a signal average of two or three different tissue classes. 
\newline
However, the development of segmentation techniques in newborn and infant brain MRI over the past decade has shown that atlas-based segmentation methodology for adult brain MR images can be applicable to child brain MRI when tuned to the specific properties of the dataset under study.  There are three major child MRI segmentation frameworks, Expectation-Maximization (EM) \citep{VanLeem1999}, Registration-based method \citep{ColZij1999} and Adaptive Label Fusion \citep{Wei2009}. 
\newline
A barrier to implementation of these methods is the dependency on a pediatric brain atlas with accurate measures of spatial boundary uncertainty for tissue classes that the early childhood dataset does not possess. Up to date, the creation of such standard atlas that would cover the entire developmental epoch still remains a significant challenge since it requires semi-automated segmentation of large datasets.
\newline
Another methodological issue with EM-based and Adaptive Label Fusion methods in applications to infant brain MRI is using global intensity statistics that requires for each tissue type to produce similar intensities in different parts of the image. 
\subsection{Handling intra-class variability}
It is important to find a strategy to cope with a highly variable signal intensity since the same tissue type having different intensities in different parts of the brain is prone to misclassification. An effective remedy to this problem was offered by Pappas \citep{Pappas1992}. The author devised an adaptive clustering algorithm that estimates the mean intensity function (representing the pure space-dependent signal) and the corresponding posterior probabilities of the classes for all pixels in a 2D image with a varying window size. In this way, adaptability of the mean intensity function from the initially global estimates of the class means to the local characteristics of the image was achieved. However, the implementation of this adaptive clustering method to a 3D image is computationally expensive due to the slow process of successive updates of mean class intensities for each pixel within the window making the extension of this technique to 3D MRI segmentation unfeasible.
\newline  
To tackle high intra-class intensity variability im EM framework, Prastawa et al. \citep{Pras2004} used a technique developed by \citep{VanLeem1999} incorporating bias field correction step into EM algorithm. The segmentation results were refined by pruning the training samples used to estimate the probability distributions of classes for each voxel by Parzen windowing \citep{Wells1996}. The successful EM classification of child brain MRI is achieved at a high computational cost associated with overcoming restrictions of the global Gaussian mixture modelling.
\newline
\citep{Toh2004} included of the partial volume effect (PVE) model \citep{Choi1991} in EM framework that estimated mixing tissue proportions within each voxel and improved detection GM, WM and the CSF in small structures and areas between sulcal CSF and the surrounding GM.
The PVE algorithm has been successfully applied to single or multi-channel adult brain scans. In this work, we investigate the performance of PVE technique in segmentation of young child's brain MRI. 
\newline
Another approach is local segmentation by brain splitting introduced by  \citep{Xue2007}. The brain was subdivided into regions with different statistical properties of WM samples using k-means clustering followed by a Voronoi tesselation. The EM segmentation performed on each Voronoi region prevented overestimation of cortical GM. However, the number of subvolumes in Xue's brain splitting algorithm is arbitrary. 
\newline
The label fusion technique handles tissue intensity variability by means of constructing a probabilistic atlas from a small cohort of newborn brain MRI  and incorporating it into the fused segmentation model as a spatial prior \citep{War2000}. This introduces dependency on a brain atlas that our MR dataset of early brain development does not possess. 
\subsection{The proposed framework}
We propose a novel local atlas-independent framework inspired by modern trends in Computer Vision such as Kernel Fisher Discriminant Analysis (KFDA) for pattern recognition and structural similarity index (SSIM; \citep{Wang2004}) for perceptual image quality evaluation. KFDA is related to kernel-based classifiers such as Support Vector Machines (SVM) introduced by \citep{Vapnik1998}. KFDA has shown competitive performance compared to other state-of-the-art classifiers such as SVM and AdaBoost \citep{ Mika1999, Ratsch1998}.
\newline
In the proposed framework, the KFDA discriminant criterion is modified by including a regularization term that penalizes intensity differences in the neighbourhood of a brain voxel. A regularized solution in the abstract high-dimensional feature space yields connected, and therefore more meaningful, spatial tissue patterns.   
\newline
We build on the global version of the multi-channel KFDA-based classification algorithm introduced in \citep{Portman2013} and further refine the algorithm to improve its performance. Namely, we propose a local approach that handles within-class intensity variations by optimal partitioning of the brain into overlapping subdomains having different average intensities. 
\newline
We aim to classify age-specific pediatric templates into two major tissue classes (GM, WM) and the CSF. The classified representative templates can then be used for the classification of a subject brain MRI of a developmental age similar to the age range of the template.      
\newline
In the absence of a ``ground truth", we assess the quality of classification via SSIM that predicts image quality as perceived by the Human Visual System (HVS) \citep{WangBovik2009} which is regarded as an optimal information extractor that seeks to identify objects in the image. As such, the HVS must automatically identify structural distortions and compensate for the nonstructural distortions (e.g., image corruption by noise). The SSIM is an objective image quality metric that simulates this functionality and computes the degree of structural similarity between reference and distorted images. It has been shown that the SSIM is well-matched to the subject ratings of image databases, and therefore, it is a good approximation to the perceived image quality  \citep{Wang2003}. In this work, the SSIM finds a new role as 
\newline
Given the objectivity and effectiveness of the SSIM we apply it for comparison of Partial Volume Estimation \citep{Toh2004} and KFDA-based classification algorithms, as well as for automatic monitoring of the quality of classification. That is, we compute the structural closeness of classified 3D brain images containing GM, WM and CSF mean intensity values with their counterparts seen in an MR image type of a highest contrast and regarded as references. In a test example of a brain template for ages 8 to 11 months shown in Figure \ref{fig1} we rely on T1w as the most informative of all MR image types.
\subsection{Paper organization}
In the following we will describe the algorithm step by step, namely, optimal brain partitioning, modified KFDA-based classification guided by SSIM, and image stitching. A brief background on KFDA is provided in \ref{A1}.
Finally, we will report on classification results for brain template for ages 8 to 11 months, compare the performance of PVE and KFDA methods and discuss possible improvements. 
\section{Method}
\label{method}

 \subsection{Optimal brain partitioning}
\label{part}

We start with a mathematical formalization of the brain splitting problem. Let $X$ be a discrete random variable (r.v.) representing the bins $X_i$ of the histogram containing $n_i$ interior brain voxels with marginal distribution $\{ p_i\}=\{\frac{n_i}{N}\}$, $Y$ be the voxel-to-voxel image partition containing $N$ voxels with uniform distribution $\{ q_j\}=\{\frac{1}{N}\}$ and $\hat{Y}$ be a random clustering over $Y$ into $2^k$ clusters containing $N_j$ voxels at the partitioning step $k$.
\newline
For each of $2^{k-1}$ subvolumes at the partitioning step $k$ (referred to as $Y$, for simplicity) we aim to find the partition $\hat{Y}=\{\hat{Y}_i\}_{i=1}^{2}$ that maximizes $MI(X,\hat{Y})$ defined by 
\begin{equation}
MI(X,\hat{Y})=-\sum_{i=1}^{2}p_i \log{p_i}+\sum_{j=1}^{2} \sum_{i=1}^{2}p_{ij}\log{p_{i | j}}.
\label{eq1}
\end{equation} 
In equation (\ref{eq1}), $\{p_i\}$ is the marginal probability distribution of the histogram bins, $\{p_{ij}\}$ is a joint probability distribution of r.v. $X$ and $\hat{Y}$ and $\{p_{i | j}\}$ is the conditional probability of the r.v. $X$ taking the value $X_i$ given that the r.v. $\hat{Y}$ is the jth cluster $\hat{Y}_j$.
\begin{eqnarray}
p_i=p(X=X_i)=\frac{n_i}{N},
\label{lab1}\\
p_{ij}=p(X=X_i)p(\hat{Y}=\hat{Y}_j | X=X_i)=\frac{| \hat{Y}_j \bigcap X_i |}{N}=\frac{n_{i j}}{N},
\label{lab2}\\
p_{i | j}=p(X=X_i |\hat{Y}=\hat{Y}_j)=\frac{| \hat{Y}_j \bigcap X_i |}{N_j}=\frac{n_{i j}}{N_j}.
\label{lab3}
\end{eqnarray}   
Here, $n_{ij}$ is the number of shared voxels between the cluster $\hat{Y}_j$ and the histogram bin $X_i$, and $N_j$ is the number of voxels in $\hat{Y}_j$.
Substituting \ref{lab2} and \ref{lab3} for $p_{ij}$ and $p_{i | j}$ into equation (\ref{eq1}) we obtain
\begin{equation}
MI(X, \hat{Y})=-\sum_{i=1}^2 \frac{n_i}{N} \log{\left(\frac{n_i}{N}\right)}+\sum_{j=1}^{2} \sum_{i=1}^{2}\frac{n_{ij}}{N} \log \left(\frac{n_{ij}}{N_j}\right).
\label{h1}
\end{equation} 
The $MI$ can be rewritten in terms of entropy functions as follows
\begin{equation}
MI(X, \hat{Y})=H(X)-H(X | \hat{Y}).
\label{mi}
\end{equation} 
When $\hat{Y}$ is a single intensity voxel $Y_j$ and $X$ takes values from a set $\{X_i\}_{i=1}^{2}$  then there is no uncertainty that this voxel belongs to one of the bins. So, $H (X | Y)$ is equal to 0 implying that 
\begin{equation}
MI(X,Y)=H(X).
\label{y}
\end{equation} 

 {\sl Brain partitioning algorithm}.

\begin{enumerate}
\item Given $2^{k-1}$ brain volumes (or nodes of a binary tree at the level $k-1$) initially partition into $2^k$ subvolumes by cutting each of the volumes into $\hat{Y}_1$ and $\hat{Y}_2$ that contain three subsequent slices and the rest of the brain slices, respectively, with sagittal, coronal and axial planes. For each direction (sagittal, coronal, axial) compute $MI$ \ref{h1} between the histogram bins $\{X_1,X_2\}$ and the initial clusters.
\item For each subvolume $Y$ and each direction, create a set of new clusters by moving the cutting plane one slice at a time (until only three slices remain in $\hat{Y}_2$). Find the clustering that maximizes $MI$ over a set of all possible clusters  $\hat{Y}=\{\hat{Y}_1,\hat{Y}_2\},\: Y_i \in Y,\: i=1,2$ in all directions.
\begin{equation*}
{\hat{Y}}^*=\underset{\{\hat{Y}_1 ,\hat{Y}_2\}}{\arg \max\:} {MI(X,\hat{Y})}.
\end{equation*}
\item Given the total number $M$ of brain subvolumes that possess $MI$ from the top level $1$ down to the current level $k$ of the binary tree compute the total ${MI}_t$ acquired in the partitioning process. $MI_t$ is the weighted sum of the $MI_i$ associated with the subdomains $i,\: i=1,...,M$. 
\begin{equation}
\begin{aligned}
MI_{t}(X,\hat{Y}^*)=\sum_{i=1}^M\frac{N_i}{N}{MI}_i & =\sum_{i=1}^M{\frac{N_i}{N}}{H_i(X)}-\sum_{i=1}^M{\frac{N_i}{N}}{H_i(X | \hat{Y}^*)}
\label{tot}
\\
& ={H_t(X)}-{H_t(X | \hat{Y})}.
\end{aligned}
\end{equation}
The weights $\frac{N_i}{N}$ are relative volumes of the image subdomains, where $N$ is the number of voxels comprising the entire 3D brain volume. 
\item Compute the Mutual Information Ratio ($MIR$) curve using equation (\ref{tot}) 
\begin{equation}
MIR(X,\hat{Y}^*)=\frac{{MI}_t(X,\hat{Y}^*)}{{MI}_t(X,Y)},  
\end{equation}
where ${MI}_t(X,Y)=H_t(X)$ due to (\ref{y}).
\item Compute signal-to-noise ratio (SNR) averaged over newly obtained $2^k$ brain subvolumes to control growth of the MIR. Construct the SNR curve with respect to the number of subvolumes $2^{i},\: i=1,2,...,k$ and normalize it to the range of MIR values.
\item Move down to the level $k$ of the binary tree and repeat the procedure from step 1 until the following stopping criterion is satisfied. \\
{\sl Stopping criterion}: Find the abscissa of the intercept between the MIR and SNR curves that provides an optimal number of subvolumes. 
\end{enumerate}
{\sl Note}: With the growing number of partitioned brain subvolumes the acquisition of information increases. Therefore, the $MIR$ is an increasing function with respect to the number of subvolumes.
As the subdomains decrease in size in the process of brain splitting, the SNR decreases. The subvolume sizes should allow samples of brain voxels large enough to distinguish between the constituent tissue classes. In order to control the brain subvolume sizes and deduce an optimal number of subvolumes we compute the SNR.  
In this way, the balance between the ability of the proposed method to classify image subdomains and the SNR is established. 
\newline
{\sl Algorithm implementation}.
\begin{figure}[h!]
\centering
\includegraphics[width=11cm, height=8cm]{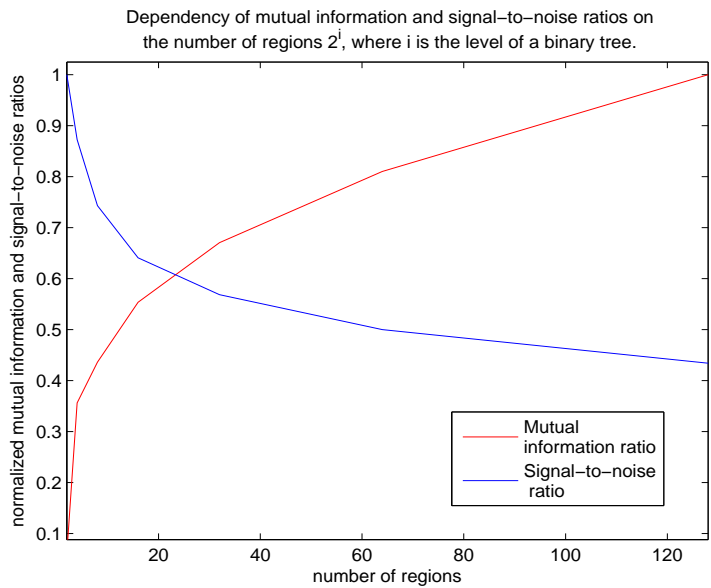}
\caption{MIR and SNR curves across 7 levels of the binary tree of the brain partitioning process in the T1w template for ages 44 to 60 months.}
\label{fig8}
\end{figure}

\begin{figure}[h!]
\centering
\includegraphics[width=5cm,height=3cm]{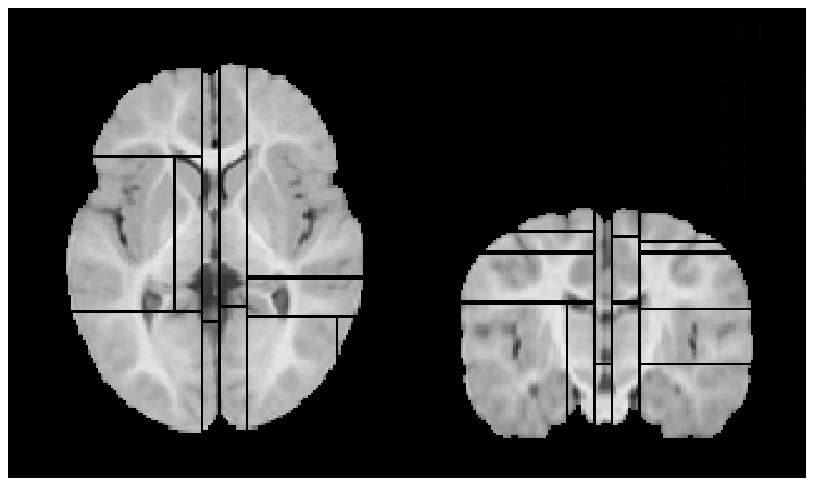}\\
\hspace{0.5cm}(a)\\
\includegraphics[width=10cm,height=6.5cm]{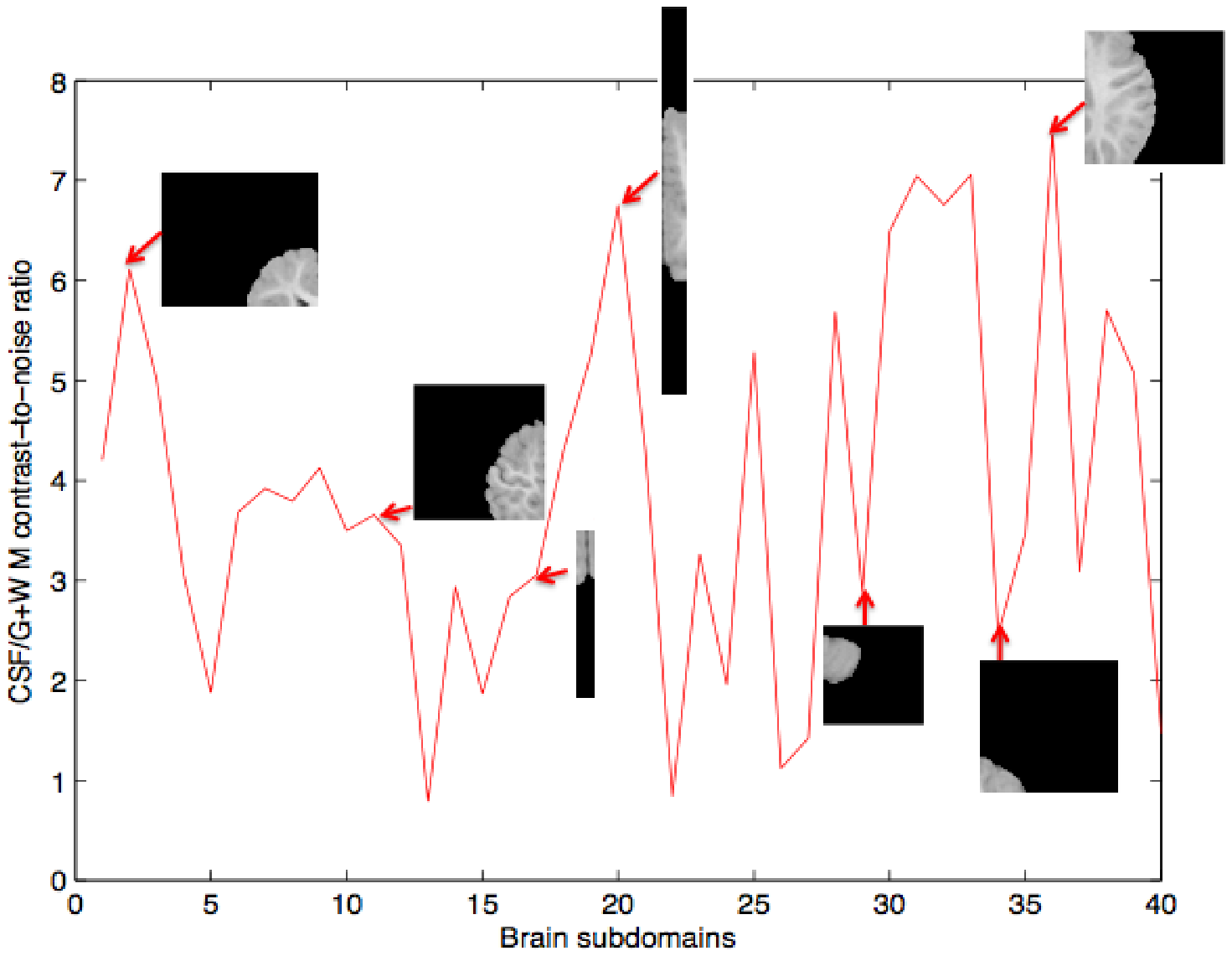}\\
\hspace{0.5cm}(b)
\caption{(a) Transversal and coronal views of the asymmetric 3D T1w brain template (ages 8 to 11 months) partitioned into 40 subdomains, (b) Local dependency of the CSF/ G+WM  contrast-to-noise ratio across the T1w template brain for ages 8 to 11 months.}
\label{cnrcsf}
\end{figure}

Figure \ref{fig8} shows the graphs of the SNR and MIR curves as functions of the number of subdomains in the T1w template for ages 44 to 60 months. The abscissa of the intercept yields the optimal number of 22 subdomains for this template. This number falls between 16 and 32 subdomains of the levels 4 and 5 of the binary tree. To determine these 22 brain subvolumes we followed a simple decision rule. We sorted the 32 subdomains in descending order according to their size and selected the first 22 subdomains with a larger size and therefore a higher SNR.
\newline 
To maintain the continuity of the classified image subdomains across their boundaries, we added two slices to each planar boundary of each subdomain thus creating an overlap of 4 slices between adjacent brain subdomains. 
\newline
Since the partitioning algorithm is intensity-based, the brain regions and their total number vary for different brain scans. Partitioning of brain MRI with a higher intensity variation yields a larger number of brain regions as demonstrated in an example below. The T1w infant brain template for ages 8 to 11 months was subdivided into 40 subvolumes (Figure \ref{cnrcsf}.a). 
Local assessment of the contrast-to-noise ratio (CNR) between the CSF and G+WM showed that CNR significantly varies across brain subdomains as seen in Figure \ref{cnrcsf}.b. It tends to be lower near the base of the brain and higher in the central part of the brain.
\newline
In a similar fashion, we can compute GM/WM CNR in image subdomains of the brain template. Local dependency of CNR graphs provides a better understanding of natural MR signal intensity variation throughout an individual brain shedding light on subvolumes with poor CSF and/or WM signal detection. Since the proposed method is kernel-based the kernel parameters (e.g., bandwidth of the Gaussian kernel) can be tuned over a certain range of values in accordance with CNR of brain subvolumes to ensure successful segmentation.

\subsection{Modified KFDA optimality criterion}
\label{math}
We reap the benefits of the KFDA approach to brain tissue classification of infant brain MRI. To mention a few, KFDA allows a simple closed form solution, improves classification by taking into account all MR signal observations (samples) in the input space, and automatically identifies PV voxels that lie near the boundary between the tissue classes. The competitiveness of the KFDA method compared to other state-of-the-art classifiers \citep{Mika1999} has motivated our exploration of KFDA to find high accuracy segmentation solutions in child brain MRI.  
\newline
We implement a binary formulation of KFDA in the feature space $\mathcal{H}$ provided in \ref{A1}. The classical kernel Fisher discriminant given by (\ref{fin}) does not assume spatial correlations between neighbouring intensity voxels in $\mathcal{H}$.To increase robustness to misclassification caused by the presence of noise in MR images we introduce a spatial regularization term that penalizes local kernel-projected intensity differences in the feature space $\mathcal{H}$. Since the graph that defines the local relationships between the brain voxels is a 3D grid with each interior node having 26 neighbours we define the neighbourhood matrix as follows
\[
H_{ij}=
\begin{cases}
1,            &\text{if $(i,j)$ is an edge;}\\
-d_{i j},   &\text{if $i=j$, the degree of vertex $i$};\\
0,            &\text{otherwise.}
\end{cases}
\]
Then 
\begin{equation}
 \vec{V}^T H \vec{V}=-\sum_{(i,j) \in E} (V_i-V_j)^2, \: \: \forall \: \vec{V} \in R^{l},
\end{equation}
where $E$ is a set of edges $\{ V_i,V_j \}$.
\newline
Let $\vec{V}=\vec{w}\cdot \phi(\vec{I})$  be the kernel projection of the input data $\vec{I}$ onto the optimal direction $\vec{w}$ in $\mathcal{H}$. $\vec{V}$ can be rewritten as $\vec{V}=\sum_{i=1}^{l} \alpha_i K(\vec{I_i},\vec{I})$ due to the expansion of $\vec{w}=\sum_{i=1}^{l} \alpha_i \Phi(\vec{I}_i)$ in $\mathcal{H}$ spanned by the mapped training samples $\phi(\vec{I}_i), \: i=1,2,...,l$.  
 We modified the KFDA optimality criterion by adding the penalty term of the form $V^T H V=\alpha^T k H k^T \alpha$, where $k$ is the kernel matrix of size $l\times l$  
 \begin{equation}
\hat{\alpha}=\arg \max_{\alpha}\left( \frac{\alpha^T M \alpha+ \lambda \alpha^T k H k^T \alpha}{\alpha^T (N+\beta I) \alpha} \right).
\label{eq}
\end{equation}
In this setup, the penalty function forces misclassified voxels closer to another class cluster centroid.\\The problem (\ref{eq}) can be reformulated as 
\begin{equation}
\boxed{\hat{\alpha}=\arg \min_{\alpha} \left(-\frac{1}{2}\left( \alpha^T M \alpha+ \lambda \alpha^T k H k^T \alpha \right) \right)\\
 \text{subject to}\:  \alpha^T (N+\beta I) \alpha=1.} 
 \label{neweq}
\end{equation}
The constrained optimization problem (\ref{neweq}) can be solved algebraically using the method of Lagrange multipliers. We constructed the Lagrangian function 
\begin{equation}
 L=-\frac{1}{2}\left( \alpha^T M \alpha+ \lambda \alpha^T k H k^T \alpha \right) +\frac{1}{2}\gamma(\alpha^T (N+\beta I) \alpha-1)
\end{equation} 
and by computing the gradient of the Lagrangian $L$ with respect to $\alpha$ we obtained an eigen-value problem
\begin{equation}
(N+\beta I)^{-1}(M+\lambda k H k ^T)\alpha=\gamma \alpha.
\label{eig}
\end{equation}
Then the solution to (\ref{eq}) is a leading eigen-vector of $(N+\beta I)^{-1} (M+\lambda k H k^T)$. 
\subsection{Objective assessment of classification quality via SSIM}
\label{ssim}
A modified version of the KFDA criterion given by (\ref{eq}) depends on a regularization coefficient $\lambda$ that influences the quality of classification. In order to find the ``best'' value of $\lambda$ we constructed a sequence of $\lambda$-values $\{\lambda_i=0.000025\cdot i\},\: i=1,...,n$ and computed the Structural Similarity Index (SSIM) \citep{Wang2002, Wang2004} between the classified and structural MR input data for each $\lambda$-value in the sequence. In this application, the SSIM quantifies the degree of structural similarity between spatial tissue patterns seen in ideal (reference) and distorted (classified) images. 
\newline
We implemented the SSIM to evaluate the performance of our classification algorithm in the absence of ground truth. We evaluated how well GM, WM and CSF boundaries are captured by our algorithm versus the boundaries visually extracted from input T1w images.
\newline
For each partitioned brain subvolume we solved a 2-class separation problem and created $n$ subvolumes with the mean T1w intensity values for the two tissue types (G+WM and CSF, and GW and WM). These are the classified $\lambda$-dependent image subdomains to be compared with T1w reference. We computed the SSIM between each classified and T1w brain slices and then found the mean SSIM taken over all slices of the brain subvolume. Thus, we obtained $n$ mean Structural Similarity Indices (MSSIM) and chose the $\lambda$ value corresponding to the largest MSSIM. In this way, we were able to automatically control the quality of classification.
\newline
Given below is a formula for the mean SSIM 
\begin{equation*}
MSSIM=\frac{1}{N} \sum_{i=1}^{N} SSIM(x_i,y_i), \: \: SSIM(x_i,y_i)=l(x_i,y_i) \cdot c(x_i,y_i) \cdot s(x_i,y_i),
\end{equation*}
\begin{figure}
\centering
\includegraphics[width=5cm,height=2.3cm]{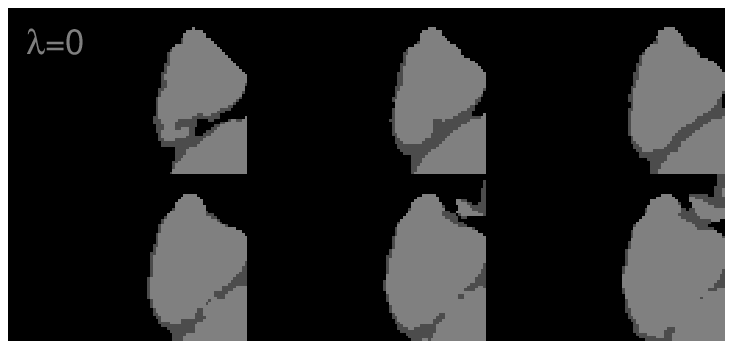}
\includegraphics[width=5cm,height=2.3cm]{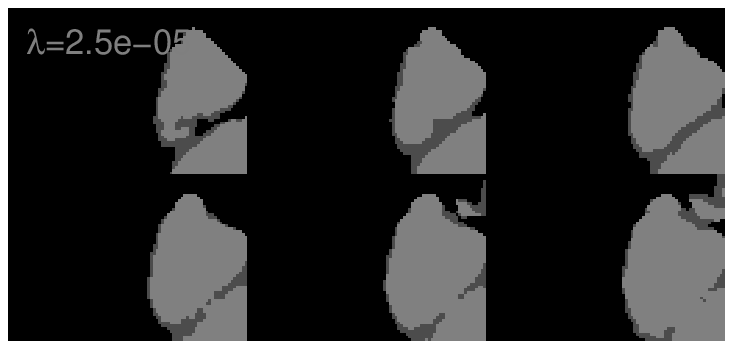}\\
\includegraphics[width=5cm,height=2.3cm]{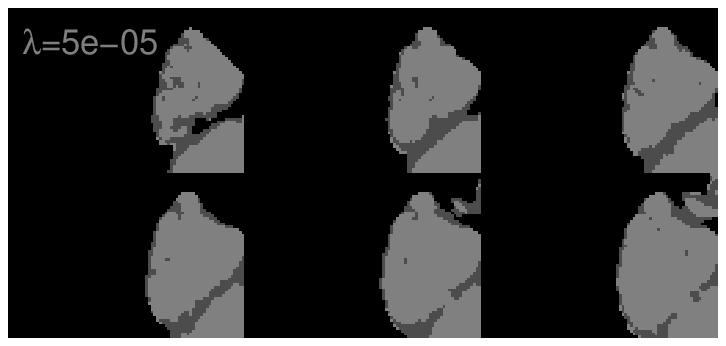}
\includegraphics[width=5cm,height=2.3cm]{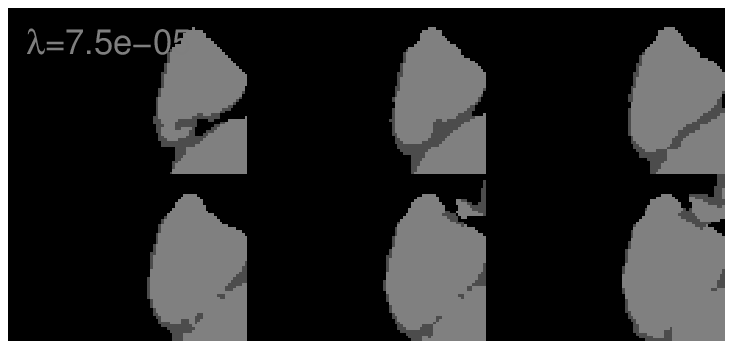}\\
\includegraphics[width=5cm,height=2.3cm]{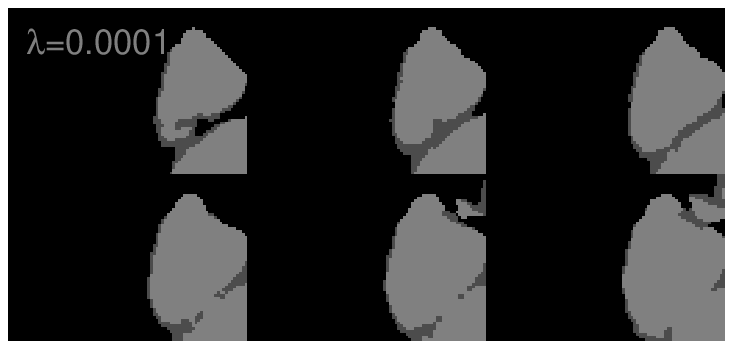}
\includegraphics[width=5cm,height=2.3cm]{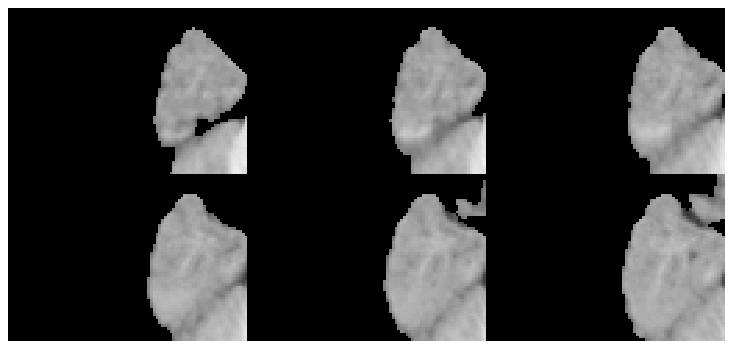}\\

\caption{Dependency of the CSF classification on $\lambda$ in the template subvolume near the base of the brain (ages 8-11 months): CSF patterns for increasing values of $\lambda$ from 0 to 0.0001 in a brain subvolume consisting of 6 slices.}
\label{sim}
\end{figure}
where  $N$ is the total number of interior brain voxels, $x_i$ and $y_i$ are local image patches\footnote[2] {a sliding window that moves across the entire brain slice pixel by pixel. For the MSSIM the background patches have been excluded.} and $l(x_i,y_i)$, $c(x_i,y_i)$, $s(x_i,y_i)$ are the luminance, contrast and structure comparison measures defined by 
\begin{eqnarray*}
l(x,y)=\frac{2\mu_x \mu_y+C_1}{\mu_x^2+\mu_y^2+C_1};\:\: c(x,y)=\frac{2\sigma_x \sigma_y +C_2}{\sigma_x^2 +\sigma_y^2+C_2};\:\:
s(x,y)=\frac{\sigma_{xy}+C_3}{\sigma_x \sigma_y+C_3}.
\end{eqnarray*}
Here, $\mu_x$($\mu_y$), $\sigma_x$($\sigma_y$) and $\sigma_{x y}$ represent the local mean, standard deviation and cross-correlation estimates of the local image patches $x$ and $y$, respectively, and $C_1, C_2, C_3$ are small constants that have been derived from the properties of the visual system \citep{Wang2002}. 
\newline
Figure \ref{sim} illustrates the dependency of the CSF pattern identified by KFDA on the value of the regularization parameter $\lambda$. The value of $\lambda=0.00005$ reveals most of the connected CSF tissue surrounding sulci, thus making the sulcus contours visible. The computed MSSIMs between the classified subdomains and the corresponding T1w template for $\lambda$-values from the sequence $\{0.000025\cdot i\}_{i=0}^{4}$ show that the CSF structure corresponding to $\lambda=0.00005$ is most similar to its counterpart in T1w reference. Therefore, the regularization term builds more of the connected CSF component by forcing initially identified G+WM voxels into the CSF class if predominant neighbouring intensities are the CSF samples.   
\subsection{KFDA implementation}
\label{kfda}
\begin{figure}
\centering
\hspace{0.05cm}
\includegraphics[width=4.7cm, height=2.9cm]{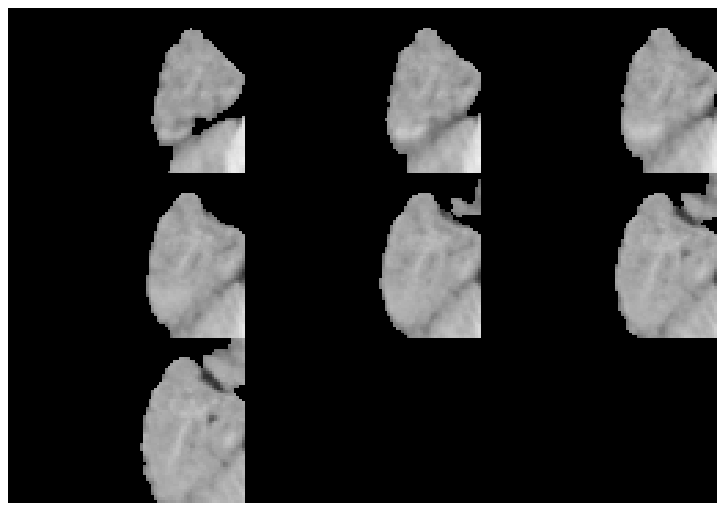}
\includegraphics[width=4.7cm,height=2.9cm]{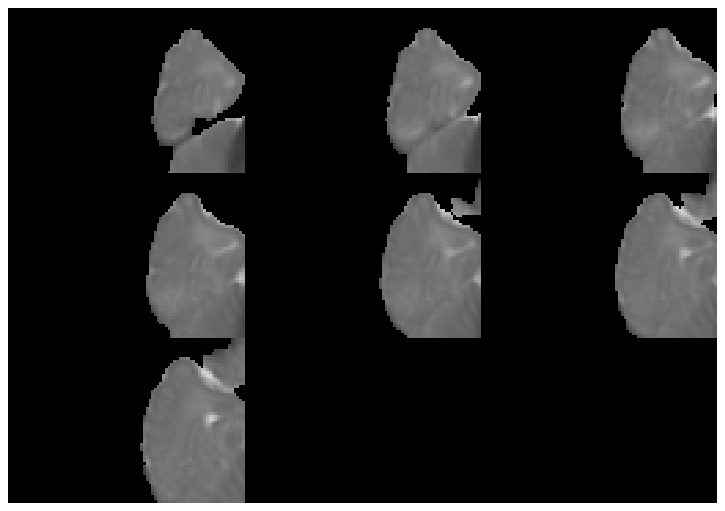}
\vspace{0.1cm}\\
\includegraphics[width=4.7cm,height=2.9cm]{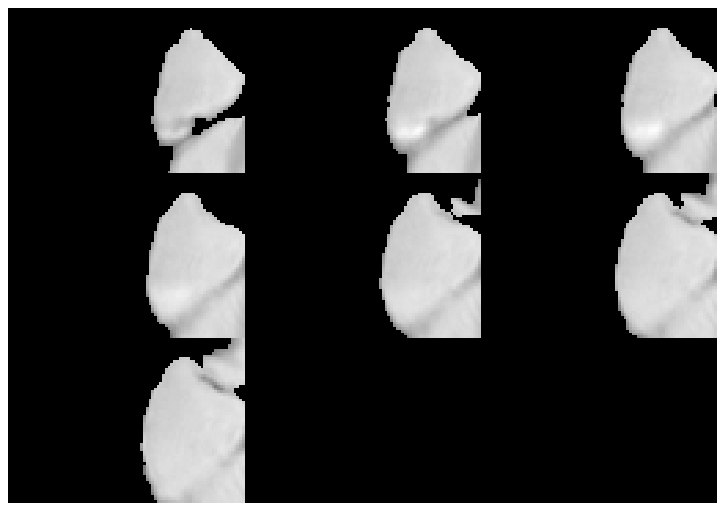}
\includegraphics[width=4.7cm,height=2.9cm]{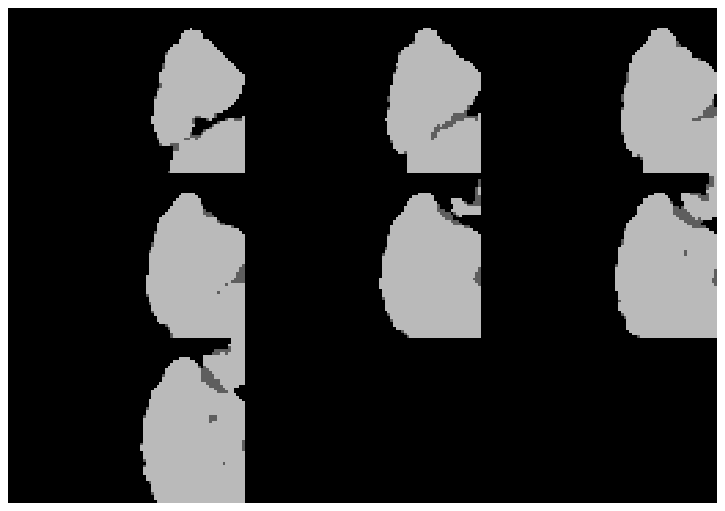}\\
(a)\\
\includegraphics[width=8cm,height=6cm]{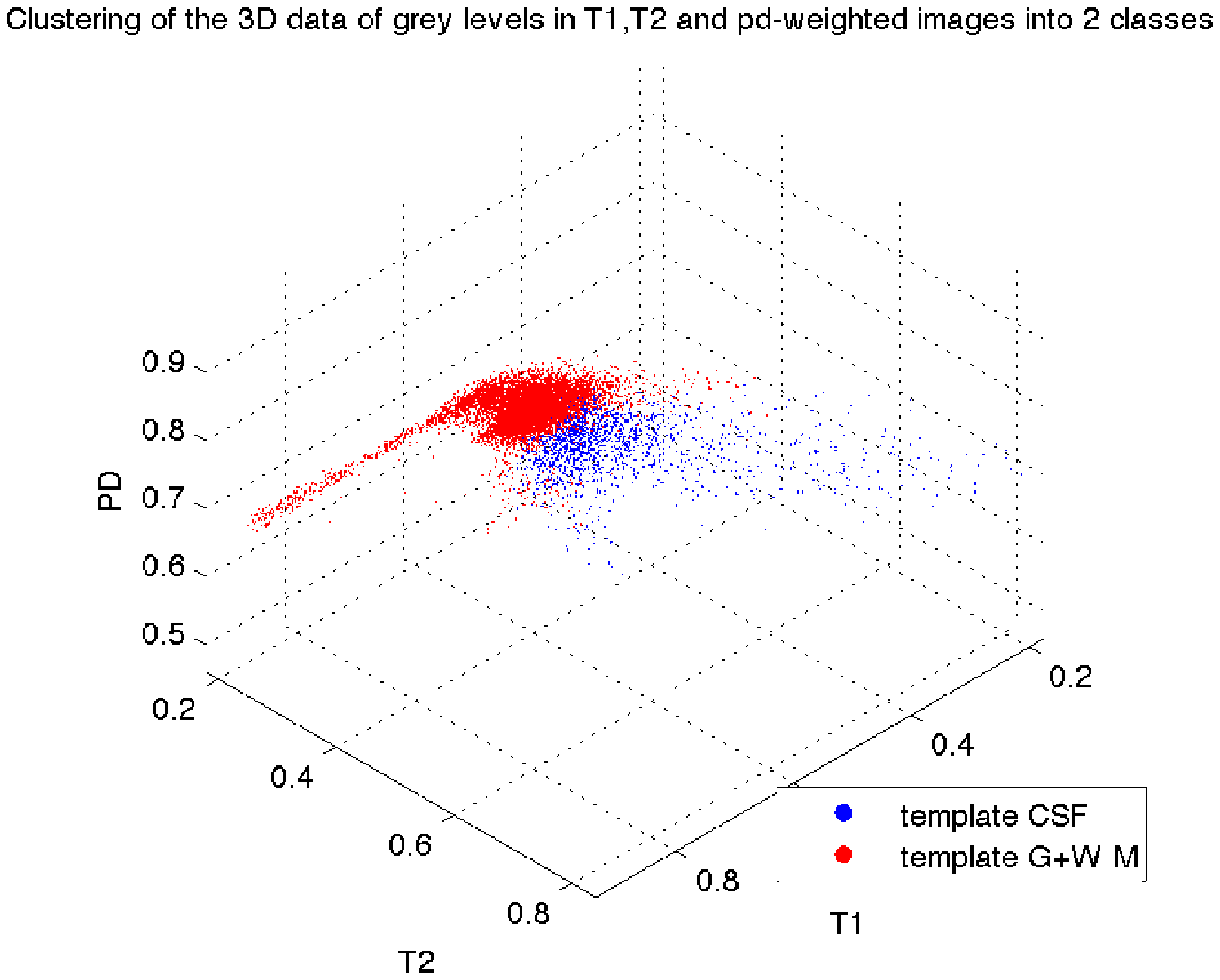}\\
(b)\\
\includegraphics[width=6.5cm,height=4.5cm]{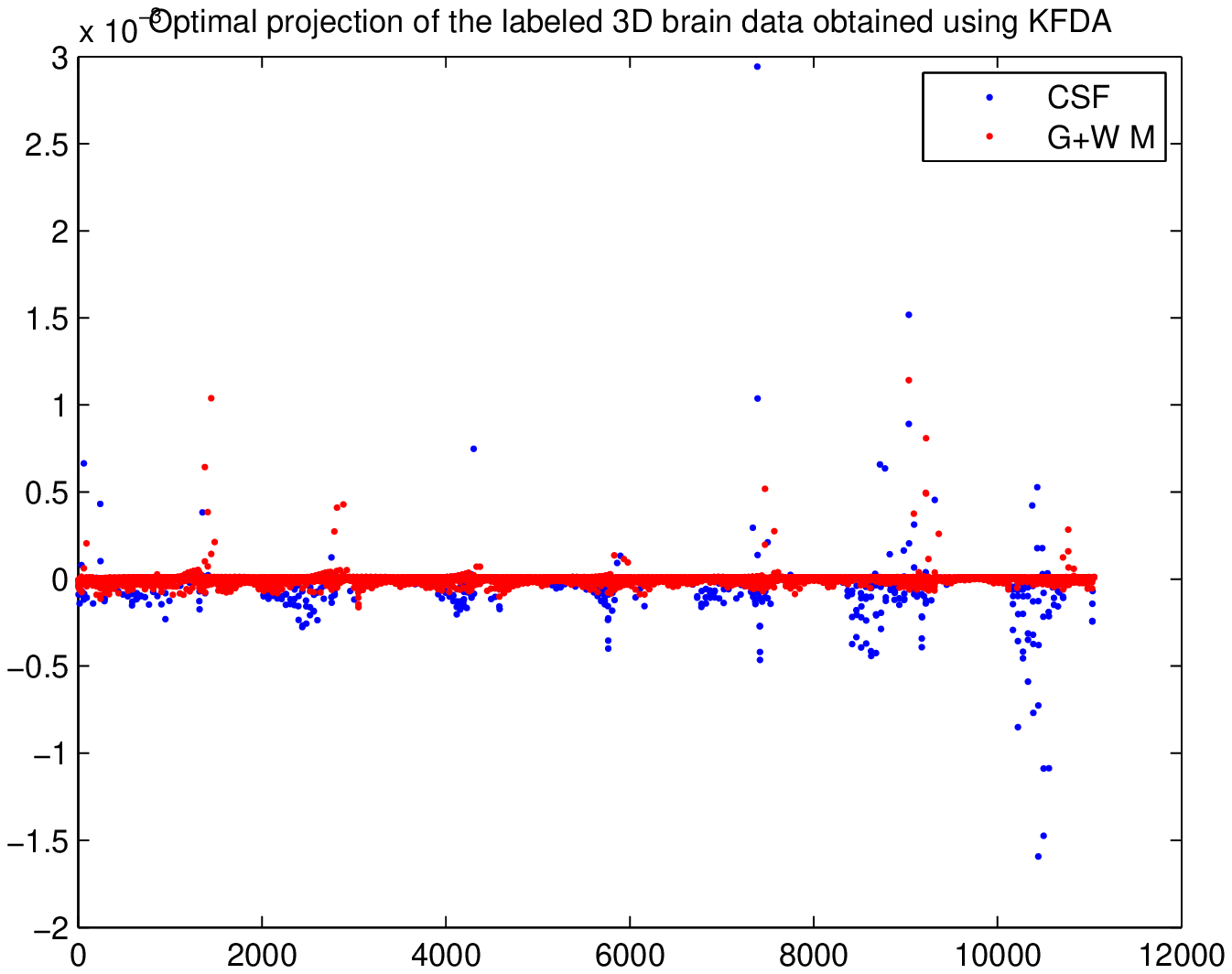}\\
(c)
\caption{(a) T1w, T2w and pdw input data from a pediatric template for ages 8 to 11 months with initial CSF and G+WM labels obtained using global PVE, (b) CSF and G+WM clusters in 3D intensity space based on the initial labeling, (c) Optimal projection of the input data given in (b) onto the vector of maximal information discrimination in the feature space. The kernel is sigmoid with $a=8$ and $b=-0.0005$.}
\label{csf}
\end{figure}

To proceed with the KFDA implementation, we solve an eigen-value problem (\ref{eig}) in a high-dimensional feature space $\mathcal{H}$ whose dimension is defined by the number of brain voxels in the image subdomain $N$. Given a vector-valued image function with labels $l=\{-1,+1\}$ for the two tissue classes (CSF and G+WM, or GM and WM) as an input
\begin{equation*}
(\vec{I}(i,j,k),l)=(t1w (i,j,k), t2w (i,j,k), pdw (i,j,k),l),
\end{equation*}
 where $1\leqslant i \leqslant M_1, \:\: 1\leqslant j \leqslant M_2,\:\: 1\leqslant k \leqslant M_3$ are voxel coordinates of the interior brain subvolume, the input data are implicitly and nonlinearly transformed to $\mathcal{H}$ 
\begin{equation*}
\vec{\phi}: \: \vec{I} \in R^3 \: \rightarrow \vec{I}^{\ast}.
\end{equation*}
Next, having chosen a kernel function $K$ and computed a leading eigen-vector $\alpha$ of the matrix given in equation (\ref{eig}) we find an optimal projection of the mapped data in $\mathcal{H}$  
\begin{equation*}
\vec{w}\cdot \vec{\phi}(\vec{I})=\sum_{m=1}^{N} {\alpha}_m K(\vec{I}_m, \vec{I})+ b.
\end{equation*}
We performed KFDA in two steps.\\
\textit{Step1: Classification into the CSF and G+WM.} 
Through extensive experimentation we determined that the sigmoidal kernel function $K(\vec{I}_m, \vec{I})=\tanh(a({\vec{I}_m}^T \cdot \vec{I})+b)$ (with $a$ and $b$ being user-specified parameters) is the best choice for delineation of the CSF. Figure \ref{csf} shows a test example of a template brain subvolume for ages 8 to 11 months that consists of 7 axial slices with initial CSF and G+WM labels obtained using global PVE. The values of the image vector-function $\vec{I}$ are plotted in the 3D intensity space shown in Figure \ref{csf}.b. Each point in the 3D input space carries specific stereotaxic coordinates $(i,j,k)$ of a brain voxel. The initial CSF and G+WM clusters are coloured in blue and red, respectively. 
\newline
An optimal projection of the input data $\vec{w}\cdot \vec{\phi}(\vec{I})$ with initial classification shown in red (G+WM) and blue (CSF) is given in Figure \ref{csf}.c. The non-linear data transformation via sigmoidal kernel preserves the sparsity of the input CSF pattern and the density of the G+WM pattern. In this Figure, the X-axis represents column-wise enumeration of the interior brain voxels from $1$ to $N$ and the Y-axis represents the projected values $\vec{w} \cdot \vec{\phi}(\vec{I}_i),\:\: 1\leqslant i \leqslant N$. When calculated with the offset $b$ they are positive for one class and negative for another. Peaks of both class clusters correspond to the voxels that lie deep inside the tissue volumes and the voxels next to the decision line $y=0$ lie close to the boundary between the CSF and G+WM.    

 \textit{Voxel categorization.} Since KFDA computes an optimal decision surface between the CSF and G+WM it easily identifies PV voxels that lie near or on the boundary between the classes.
\begin{figure}[h]
\centering
\includegraphics[width=12cm,height=4cm]{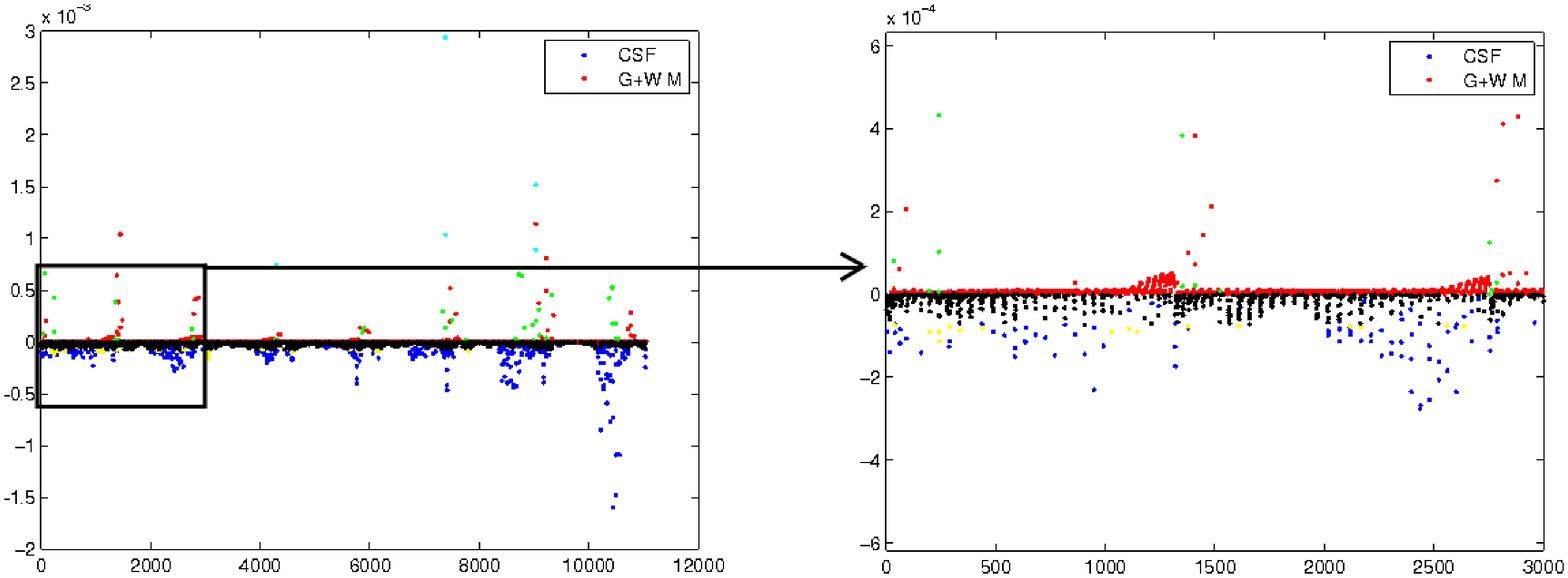}\\
(a)\\
\vspace{0.3cm}
\includegraphics[width=5cm,height=3.2cm]{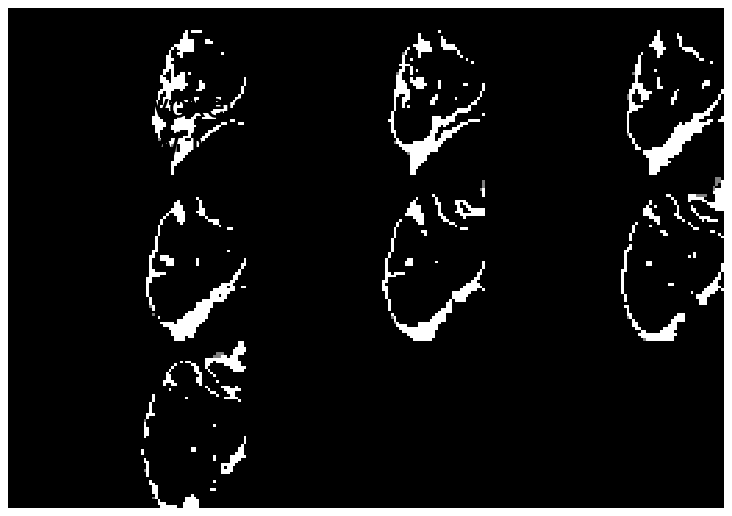}\\
(b)
\caption{(a) Voxel categorization into outliers (cyan for CSF and yellow for G+WM), overlapping set (black for G+WM and green for CSF) and tissue prototypes (red for G+WM and blue for CSF) in the feature space and their close-up view, (b) Overlapping set in the stereotaxic space.}
\label{csf3}
\end{figure}

A close-up look at the distribution of the kernel-projected data reveals overlapping CSF and G+W voxels and class outliers (Figure \ref{csf3}.a). We identifed the following voxel categories,
\begin{itemize}
\item G+WM overlapping voxels (in the negative CSF range) in black,
\item CSF overlapping voxels (in the positive G+WM range) in green,
\item CSF outliers in cyan,
\item G+WM outliers in yellow,
\item CSF and G+WM tissue prototypes in blue and red, respectively. 
\end{itemize}

A visualization of the overlapping set (coloured in black and green) in the stereotaxic space given in Figure \ref{csf3}.b shows that the overlapping voxels are located within the boundary regions between G+WM and the CSF. We recognize a particularly problematic brain area around the sulcus where the CSF is usually poorly detected due to the presence of PVE. Therefore, it is reasonable to assume that the overlapping voxels contain a mix of both tissue classes. The outliers located away from the decision boundary and their initial class centroids are most likely to change their initial membership and the rest of the brain voxels constitute tissue prototypes that truly represent CSF and G+WM tissues.

Voxel categorization is useful since we can use the predictive power of KFDA to assign the most likely class membership to overlapping voxels (containing PVE). For this purpose, we treat the overlapping voxels as a testing set and tissue prototypes as a training set. For the classification of the overlapping set we implemented \textit{$k$}-nearest neighbours(KNN) classifier. It determines the class membership based on the class majority rule in the vicinity of each voxel defined by Euclidean distances to $k$ nearest tissue prototypes in $\mathcal{H}$. The outliers comprise a separate testing set and they are classified using Mahalanobis distance. 
 \newline
It is tempting to classify all kernel-projected samples into 2 classes based on their sign (positive or negative). However, this usually leads to an overestimate of the CSF. To monitor the quality of classification produced by predictive labelling of the testing set we devised the following SSIM-guided classification procedure. 
\newline
\textit{An algorithm for SSIM-guided computation of the decision surface.}  
\begin{enumerate}
\item  Classify the outlier set using Mahalanobis distance
\item Compute the resulting classified image with tissue class means and the ${MSSIM}_{Mahal}$
\item Classify the overlapping set using KNN classifier with different values of the number of neighbours $k$
\item For each value of the parameter $k$, compute the resulting classified image and the ${MSSIM}_{knn}$. Choose the classification that corresponds to the maximal $MSSIM$
\item Compare ${MSSIM}_{Mahal}$ and ${MSSIM}_{knn}$  and return the classification that corresponds to a larger value.
\end{enumerate}     
This algorithm is general enough to handle a variety of class intensity distributions. For a test example provided by Figure \ref{csf} there are only a few class outliers and their assignment to different classes will not make a visible difference to the initial classification. Due to a large overlapping set, application of the KNN classifier contributes most to the final KFDA classification of the test example. However, there are cases when there is a large number of the class outliers and very few overlapping voxels. In this case, Mahalanobis classification of the outlier set suffices to apply.    
\begin{figure}[h]
\centering
\includegraphics[width=6cm,height=4.7cm]{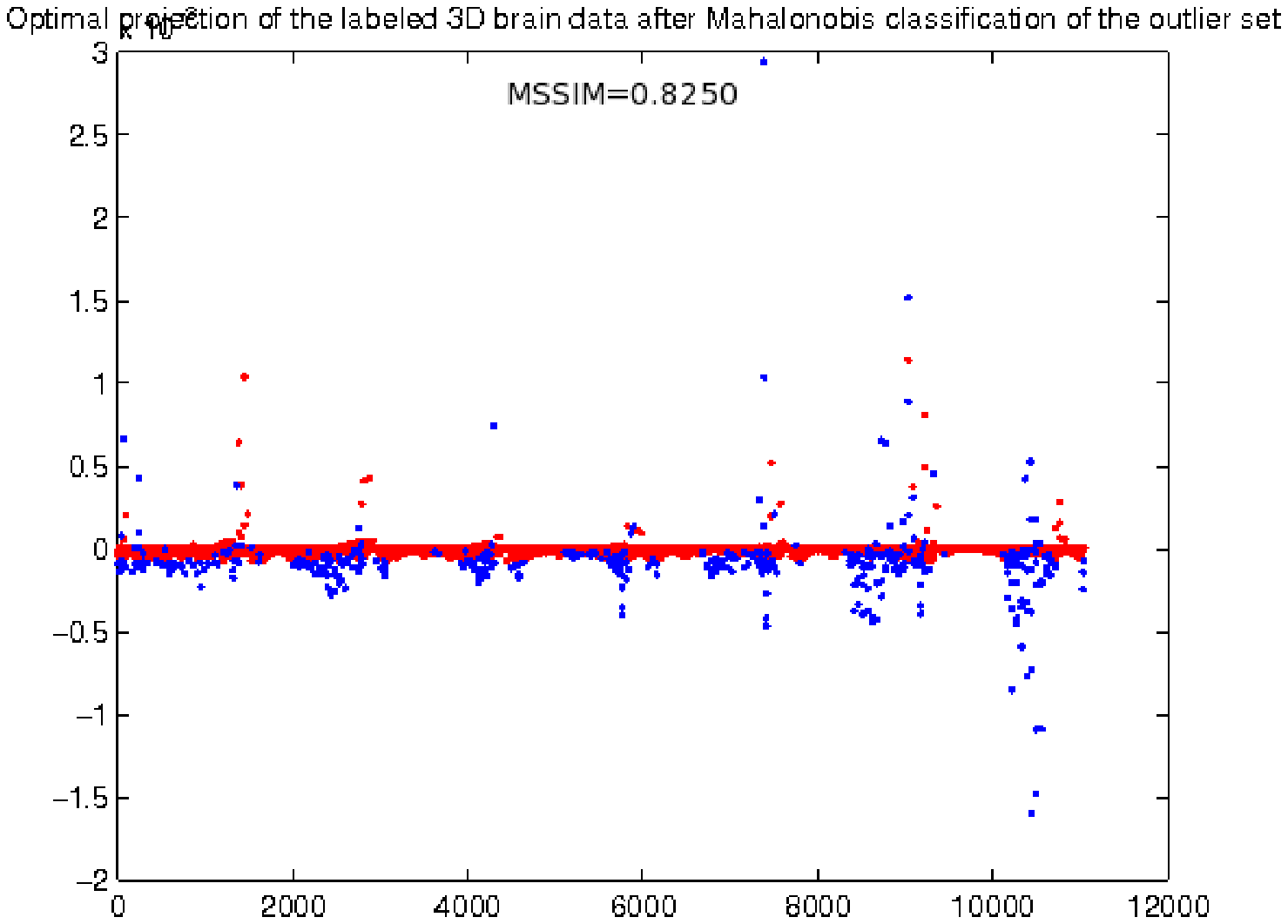}
\includegraphics[width=6cm,height=4.7cm]{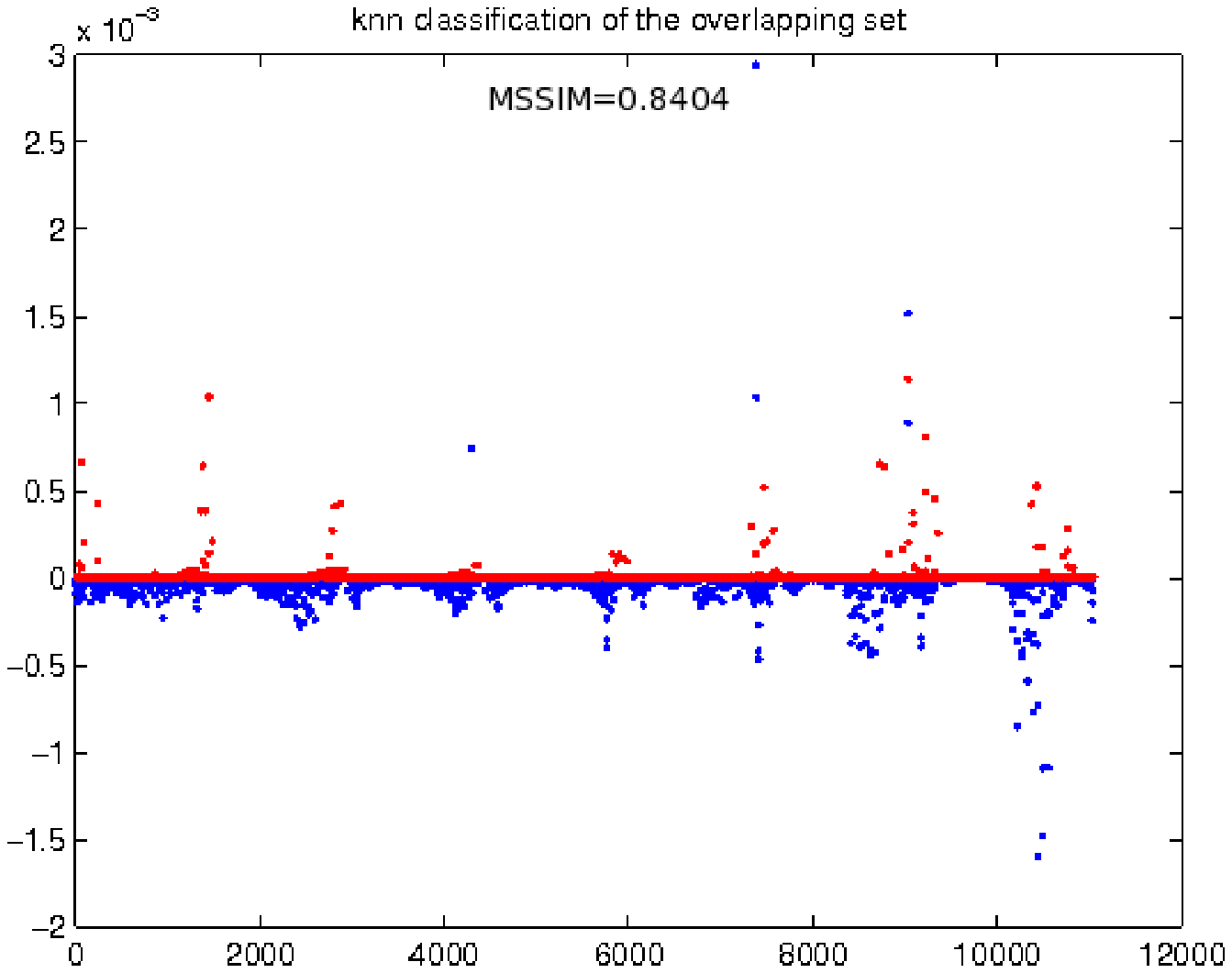}\\
\hspace{0.5cm}(a) \hspace{5.6cm} (b) \vspace{0.2cm}\\
\includegraphics[width=7cm,height=5cm]{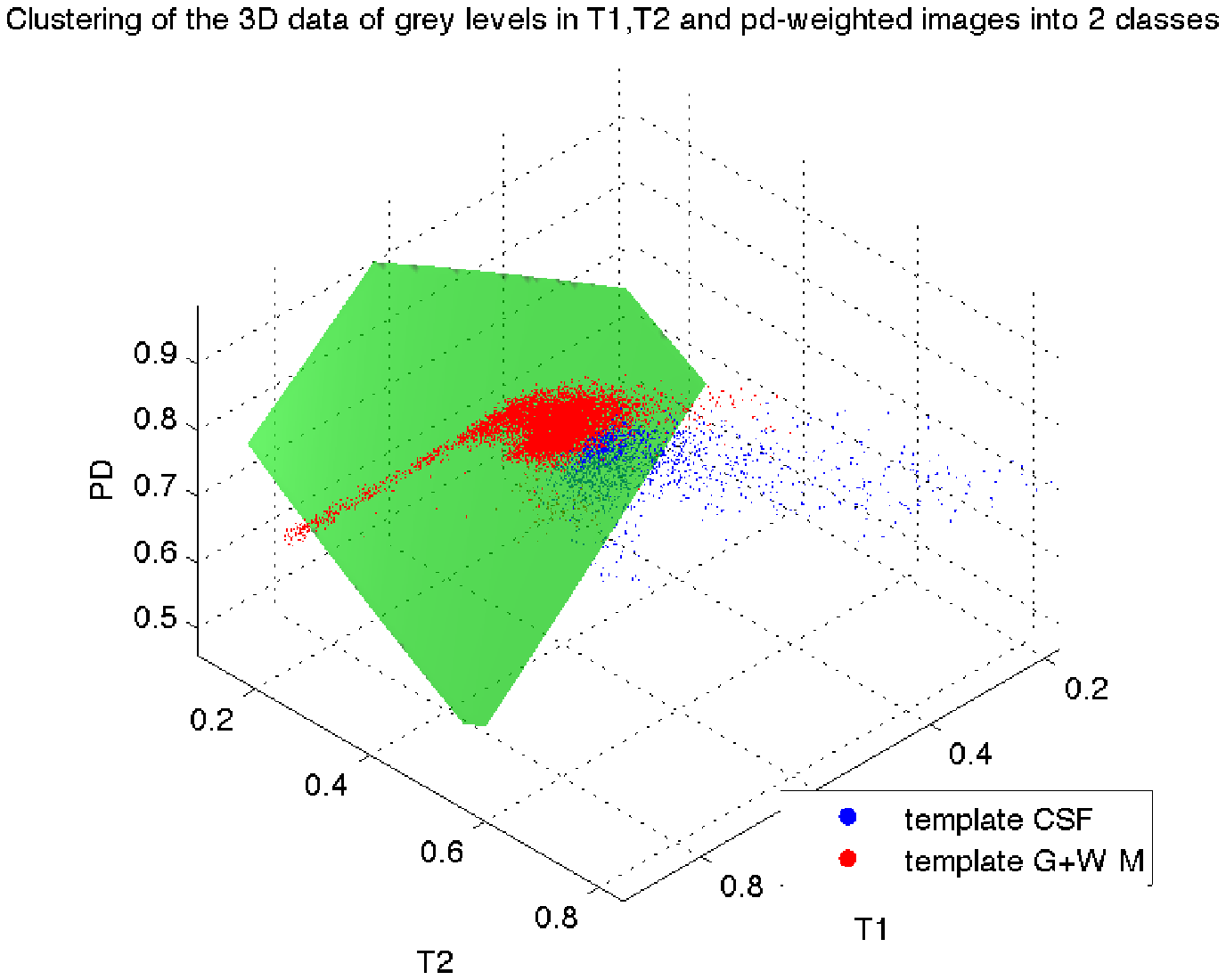}
\includegraphics[width=4.3cm, height=2.6cm]{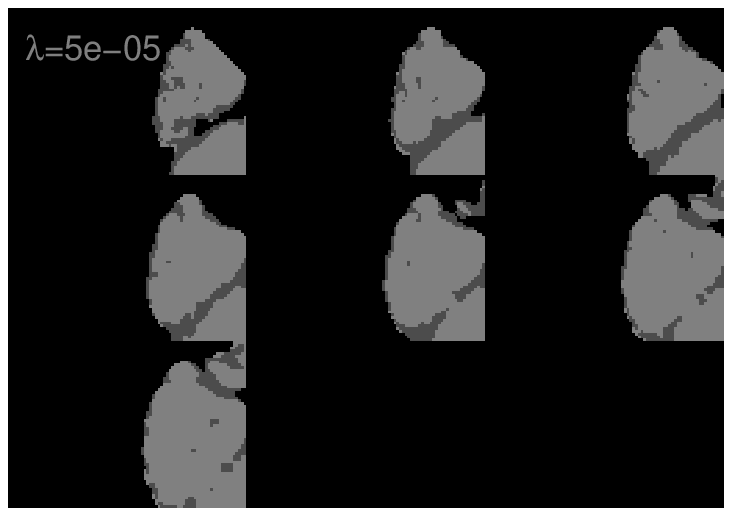}\\
\hspace{0.5cm}(c) \hspace{6cm}(d)
\caption{(a) Mahalanobis classification of the outlier set, (b) KNN classification of the overlapping set, (c) decision surface delineating the CSF cluster in the input space, (d)classification in the stereotaxic space corresponding to a larger $MSSIM$ value $MSSIM_{knn}=0.8404$.}
\label{res2}
\end{figure} 
Figures \ref{res2}.(a-b) demonstrate steps 1-4 of the algorithm, and Figure \ref{res2}.c shows the separating surface in the input space corresponding to the KNN classification with an optimal value of the parameter $k$ shown in Figure \ref{res2}.b. The discriminating boundary between the CSF and G+WM clusters is a plane. The CSF cluster identified by the SSIM-guided algorithm contains 1917 voxels, a significant increase over the initial CSF volume consisting of 515 voxels. It is shown in stereotaxic coordinates in Figure \ref{res2}.d.  
\newline
\textit{Step2: Classification into GM and WM.}
\newline
Having delineated the CSF we classify G+WM into GM and WM. Experiments with various kernel functions showed that the Gaussian radial basis function $K(\vec{I}_i, \vec{I})=\exp\left(-\frac{(\vec{I}_i-\vec{I})^T (\vec{I}_i-\vec{I})}{2\sigma^2}\right)$ is the best choice to model the non-linear structure of WM and GM clusters. The G+WM of the test example with initial GM and WM labels is given in Figures \ref{gw}.(a-b).The initial classification represents a significant underestimate of WM. The small WM structure in the temporal lobe of low signal intensity is visible but not detected by the PVE method.
\begin{figure}[t!]
\centering
\includegraphics[width=4.8cm,height=3cm]{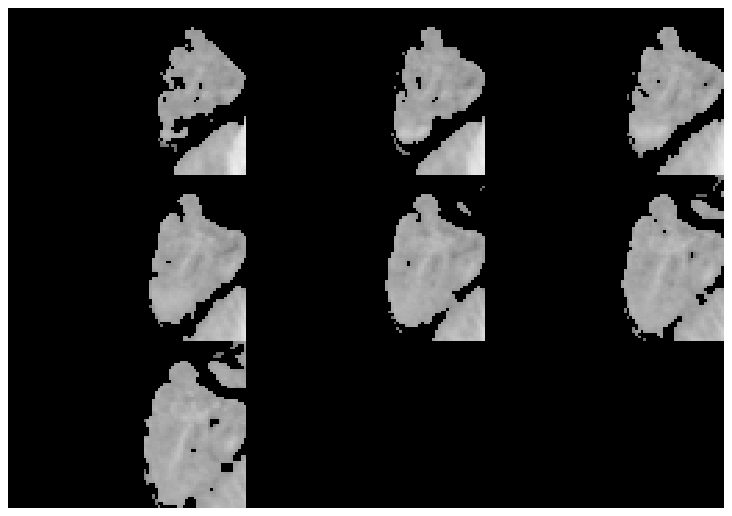}\hspace{1.3cm}
\includegraphics[width=4.8cm,height=3cm]{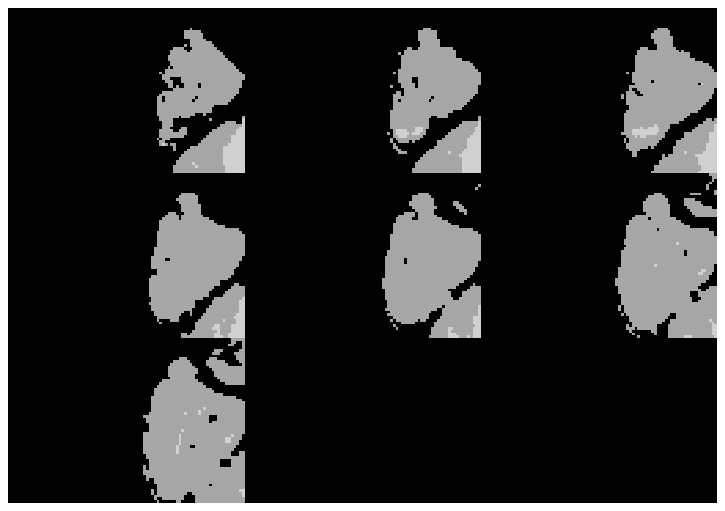}\\
(a)\hspace{6cm}(b)\\
\includegraphics[width=5.5cm,height=4.5cm]{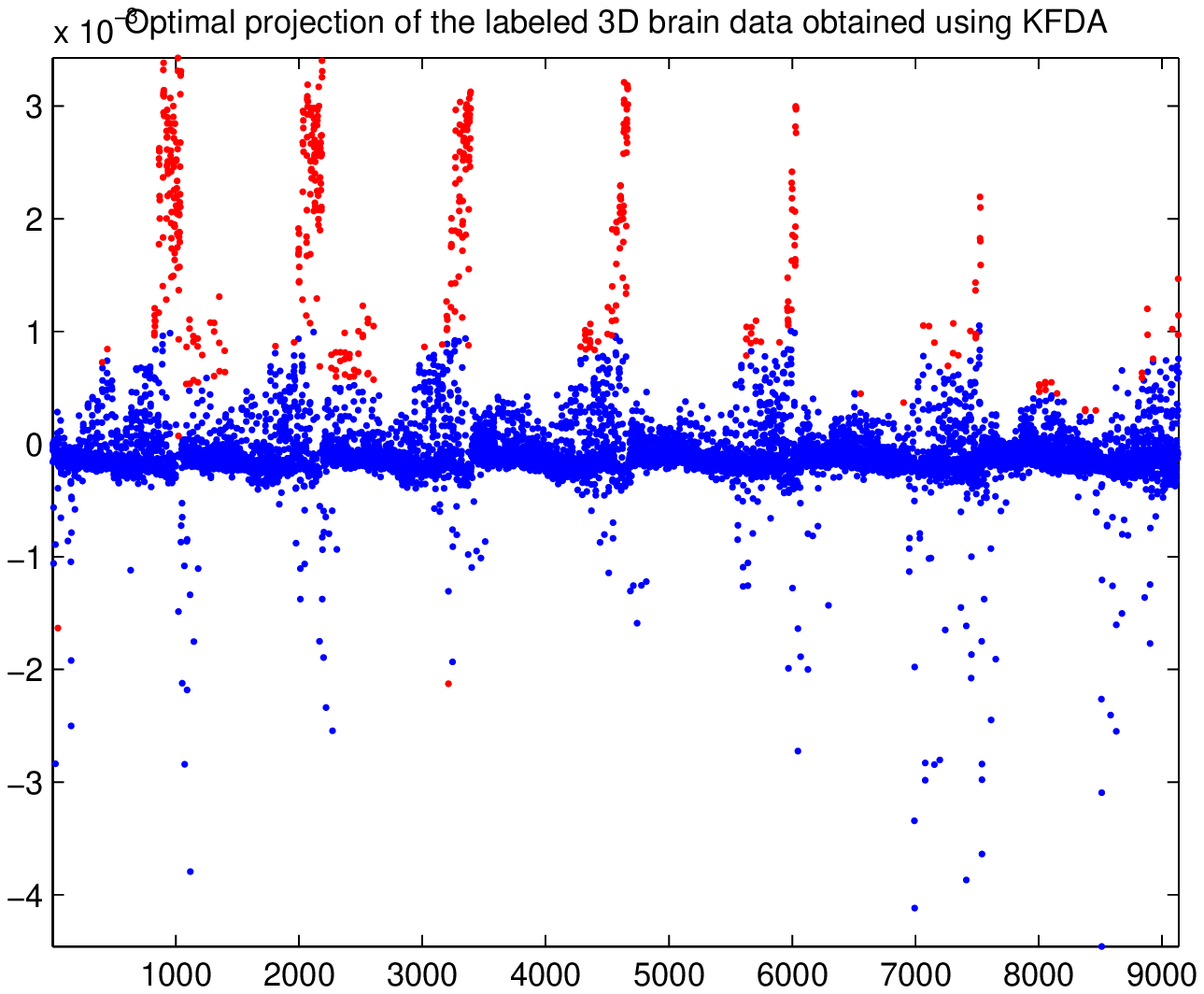}\hspace{1cm}
\includegraphics[width=5.5cm,height=4.5cm]{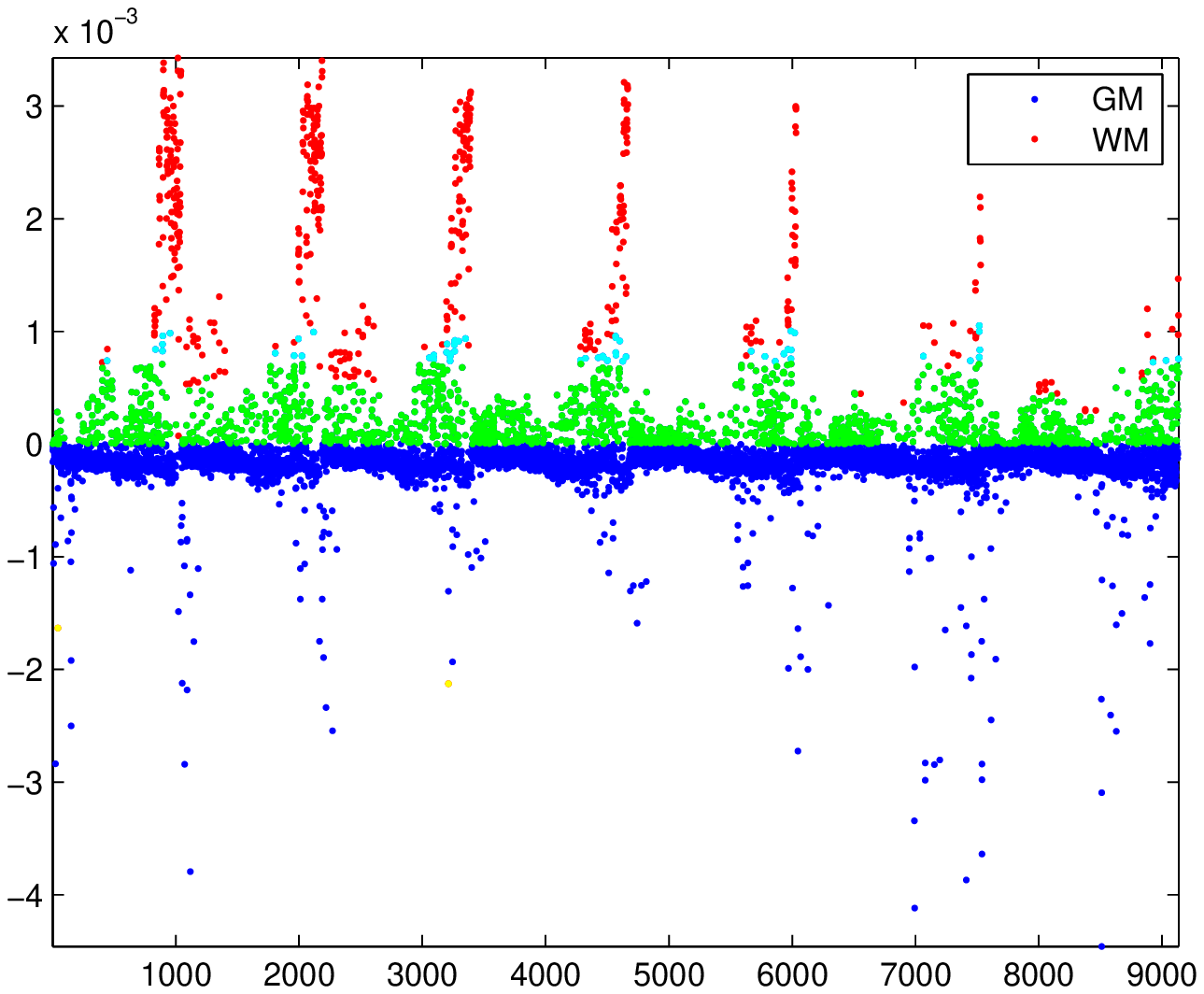}\\
(c)\hspace{6cm}(d)\\
\includegraphics[width=5.5cm,height=4.5cm]{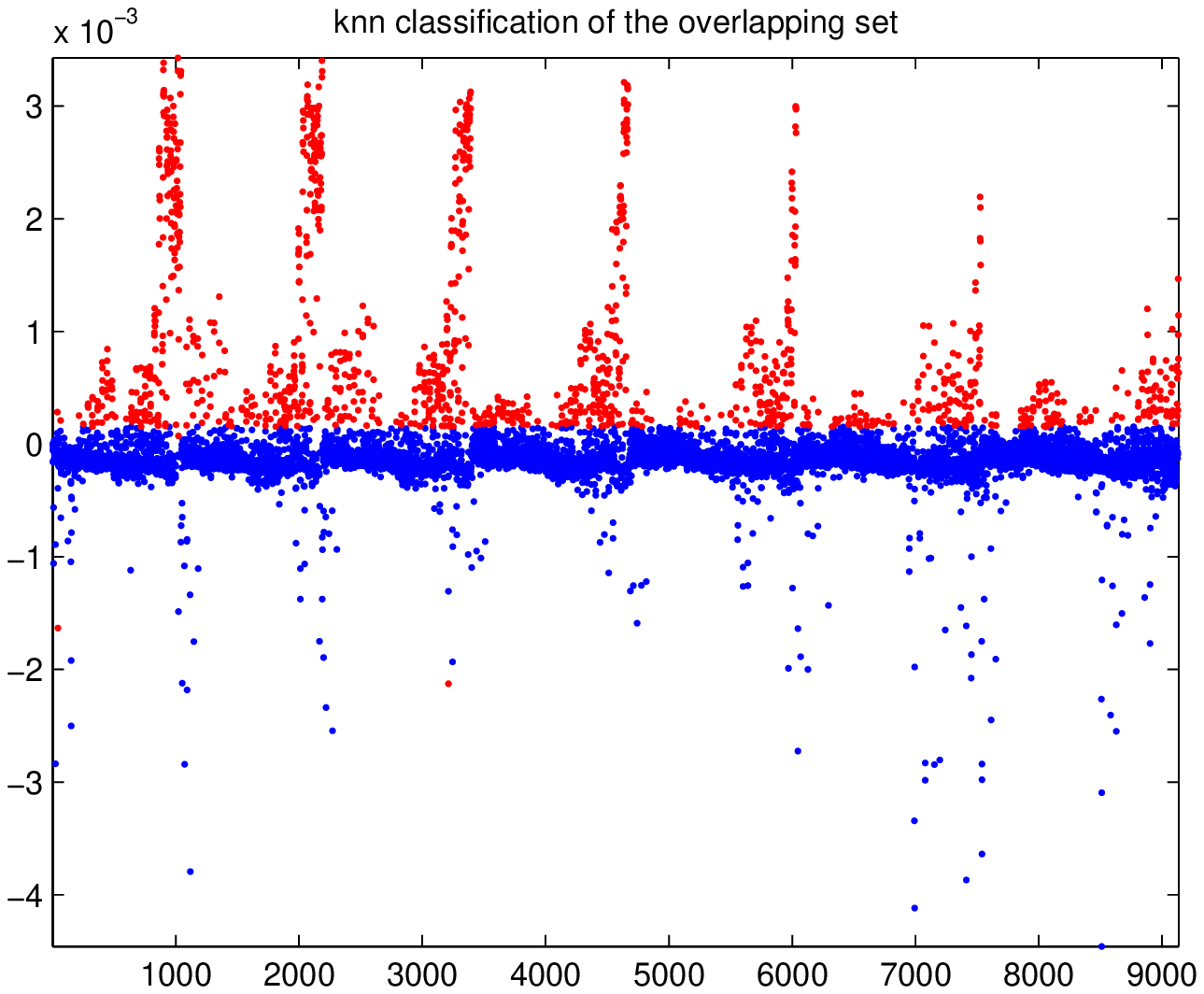}
\includegraphics[width=6.5cm,height=5cm]{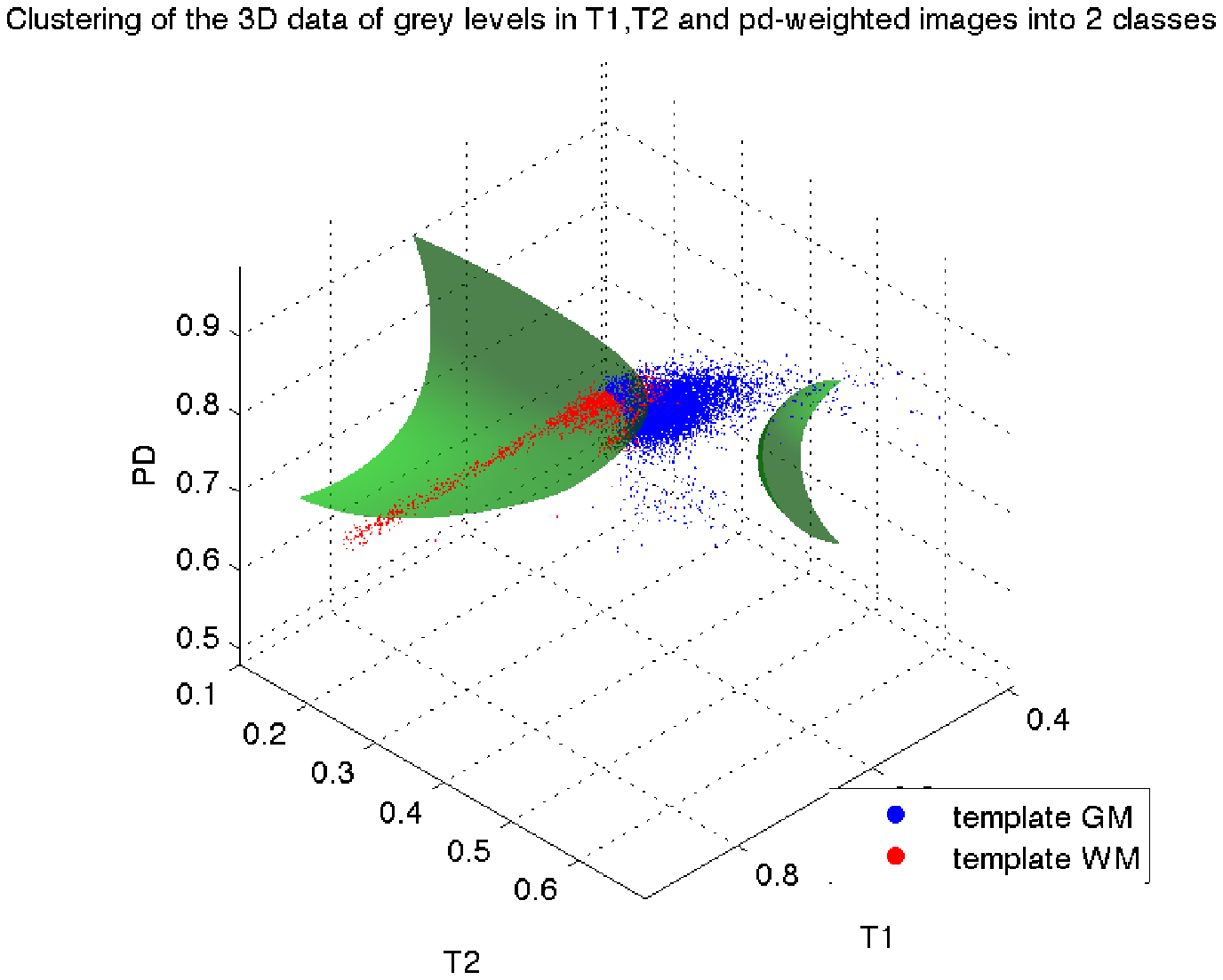}\\
(e)\hspace{6cm}(f)\\
\caption{(a-b) G+WM cluster in the T1w template brain subvolume (Figure \ref{csf}.a) with initial GM and WM labels obtained by PVE, (c) Non-linear mapping and optimal projection of the labeled G+WM input data using a Gaussian RBF with $\sigma=0.5$, (d) Voxel categories: GM (cyan) and WM (yellow) outliers and GM overlapping voxels (green), (e) GM (blue) and WM (red) clusters in the feature space, (f) Decision surface in the input space.}
\label{gw}
\end{figure}  
We performed KFDA of the labeled G+WM input data and displayed the projected GM and WM distributions in $\mathcal{H}$ as shown in Figure \ref{gw}.c. Here, WM and GM peaks correspond to the voxels located deep inside the GM and WM. Unlike the CSF and G+WM case, GM and WM distributions contain only GM overlapping voxels that fall in the positive range of WM and class outliers (Figure \ref{gw}.d). The tissue prototypes that form the training set are color-coded in red for WM and blue for GM. 
\newline
We then applied the algorithm for SSIM-guided computation of the decision surface and identified 819 WM voxels in the feature space in addition to the initial 482 WM voxels shown in red in Figure \ref{gw}.e. The corresponding non-linear decision surface that attempts to mimic the shape of the GM cluster in the input space is displayed in Figure \ref{gw}.f. Figure \ref{gw}.f also demonstrates the importance of providing multichannel data for the input. This test example possesses a wide range of T1w WM intensities and a narrow range of T2w WM intensities. Without the T2w imaging data it would not be possible to detect myelinated WM with a lower signal intensity.
SSIM comparison of KFDA and PVE classification results shows that KFDA yields spatial tissue patterns that are structurally closer to their counterparts in T1w reference subdomain (Figure \ref{gw2}). 
\begin{figure}
\centering
\includegraphics[width=12cm,height=2.6cm]{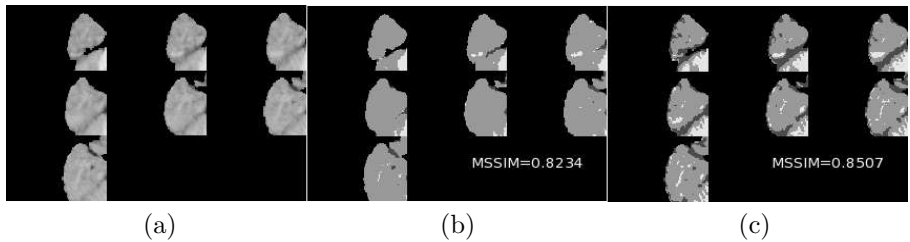}
\hspace{0.5cm}(a)\hspace{3.5cm}(b)\hspace{3.5cm}(c)\\
\caption{(a) T1w template subdomain and its (b) global PVE classification, (c) KFDA-based classification into GM, WM and CSF.}
\label{gw2}
\end{figure} 

\subsection{Stitching of brain subdomains}
\label{stitch}
We applied a Simulated Annealing technique to stitch the local classifications together into a cohesive global picture of the classified brain. First of all, for each brain slice we collected the constituent subimages. Each pair of overlapping subimages contained a joint image region $I_s$ of size $4 \times ncols$ or $nrows \times 4$ that is to be optimally estimated from two sets of observations $I_l$, $I_r$ or $I_{up}$, $I_{low}$ as illustrated in Figures \ref{st}.a-b. 

\begin{figure}[h!]
\centering
\includegraphics[width=6cm,height=5.5cm]{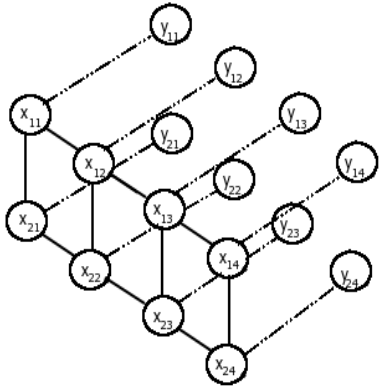}
\caption{Markov Random Field image model.}
\label{mrf}
\end{figure}

\begin{figure}[h!]
\centering
\includegraphics[width=5cm,height=4.5cm]{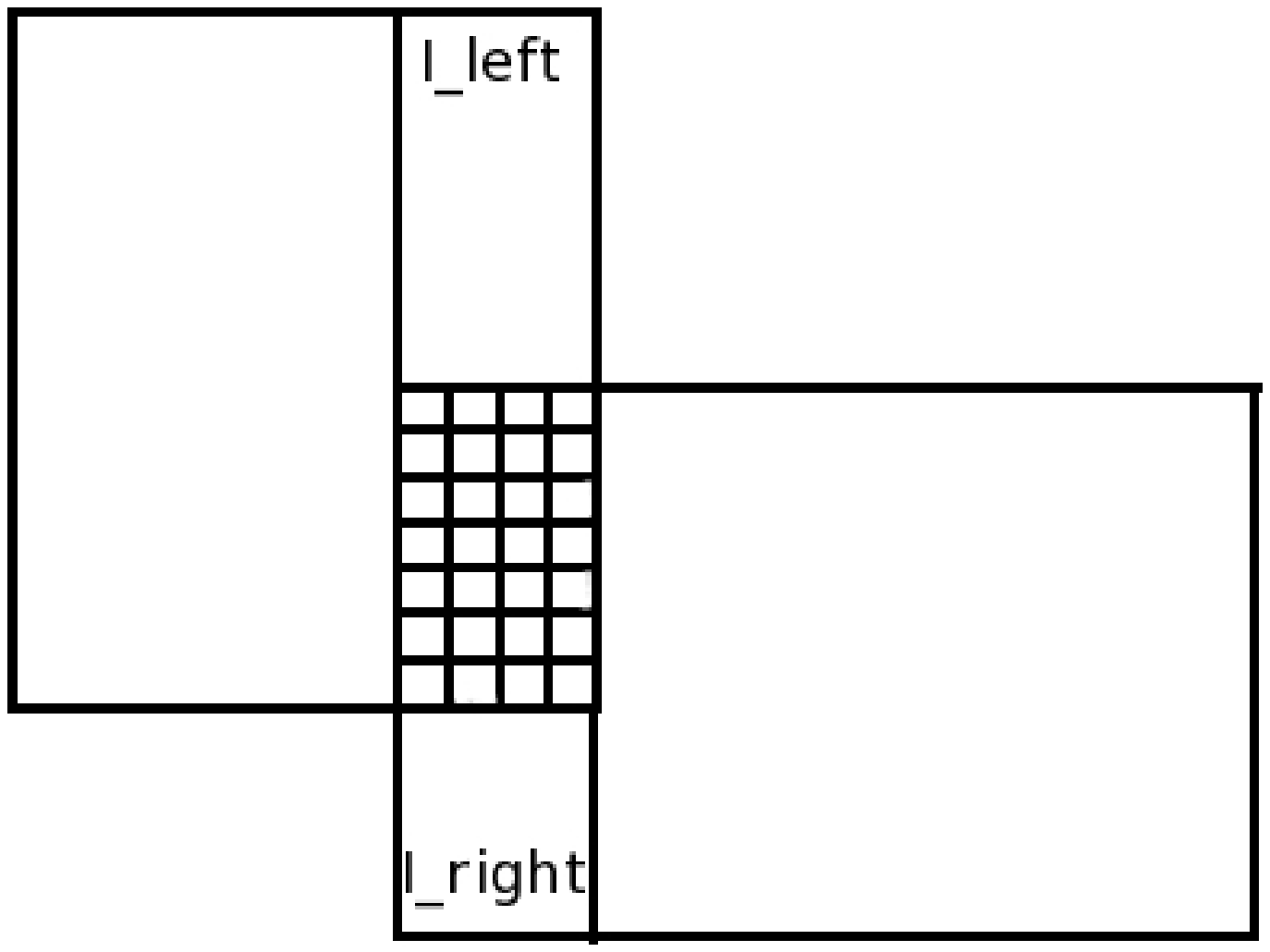}
\includegraphics[width=5cm,height=4.5cm]{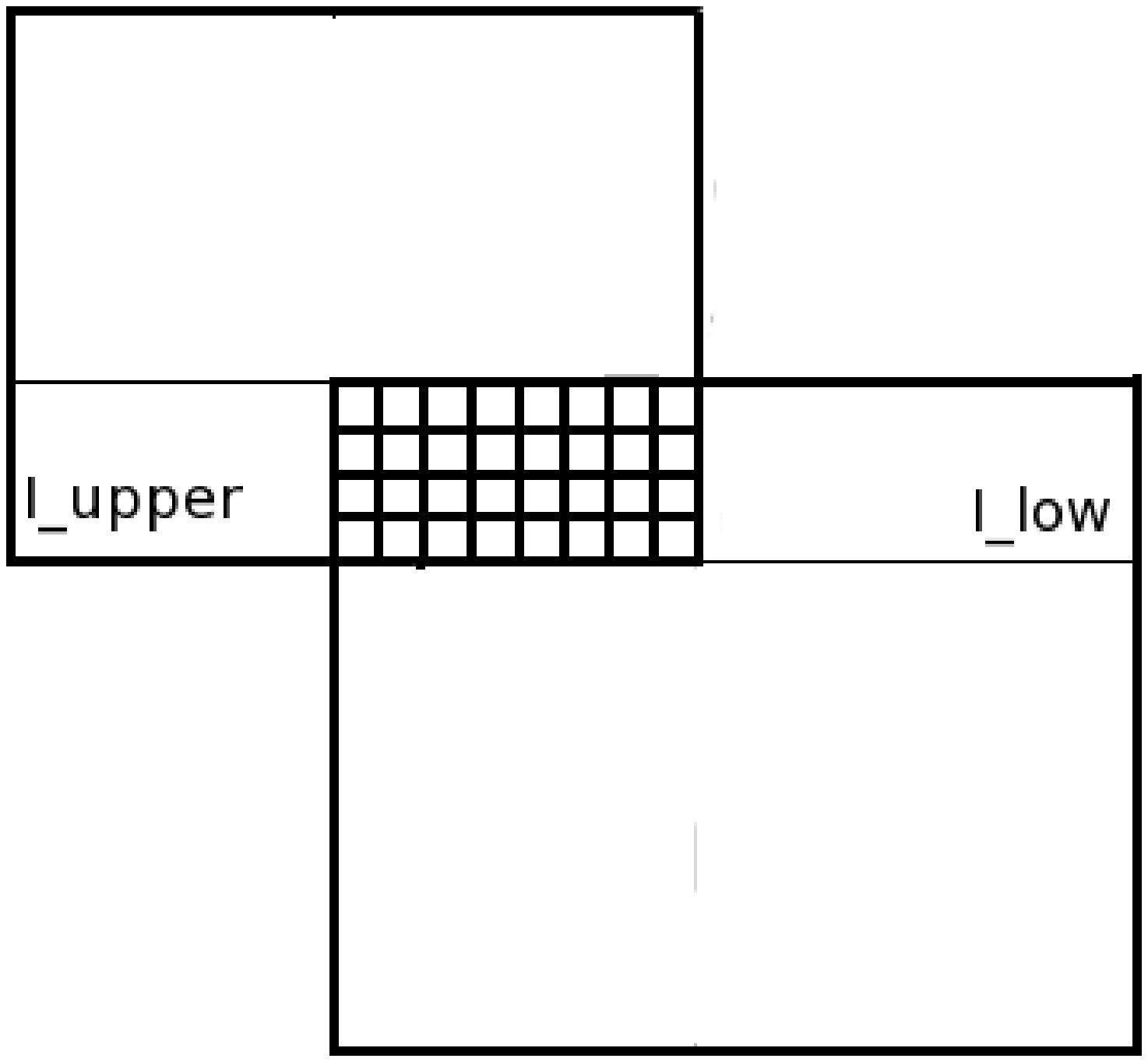}\\
(a) \hspace{3.2cm} (b) \vspace{0.3cm}\\
\includegraphics[width=9cm,height=3.8cm]{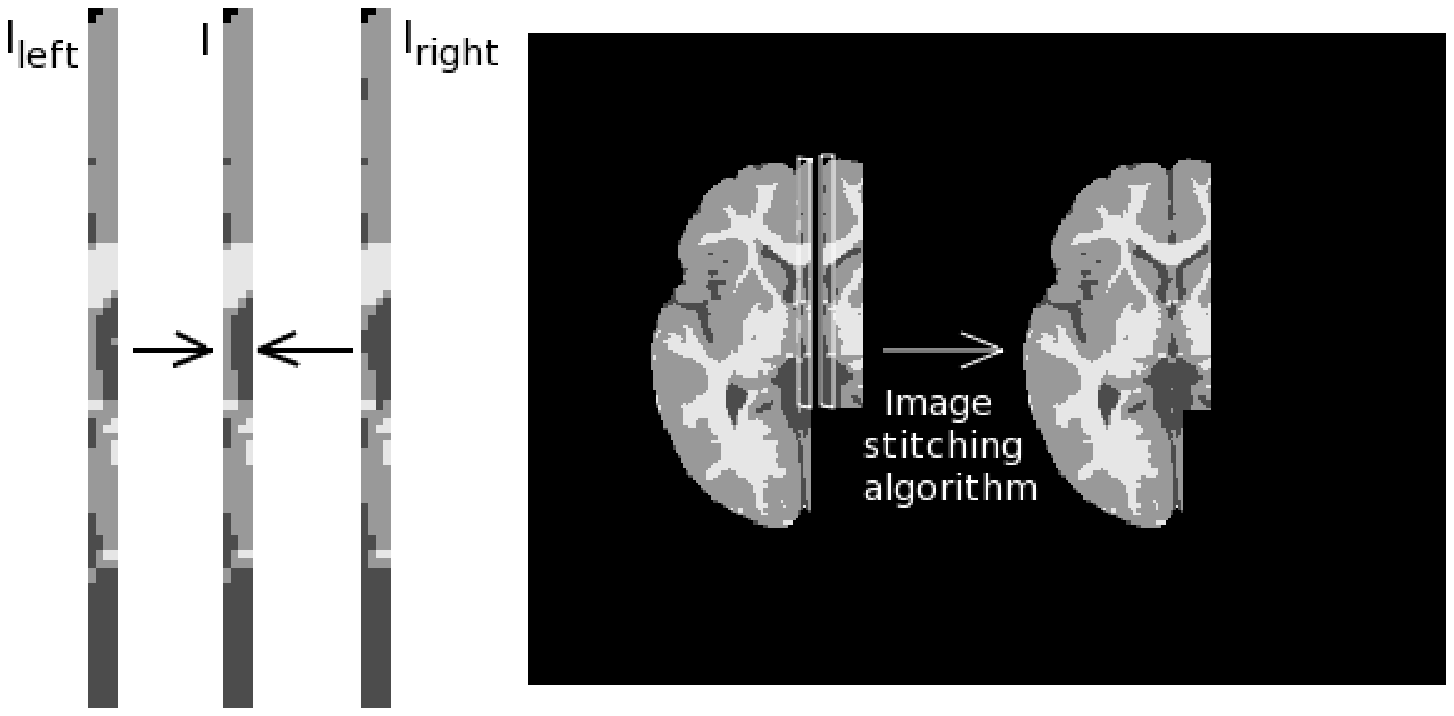}\\
(c) \vspace{0.3cm}\\
\includegraphics[width=9cm,height=6.5cm]{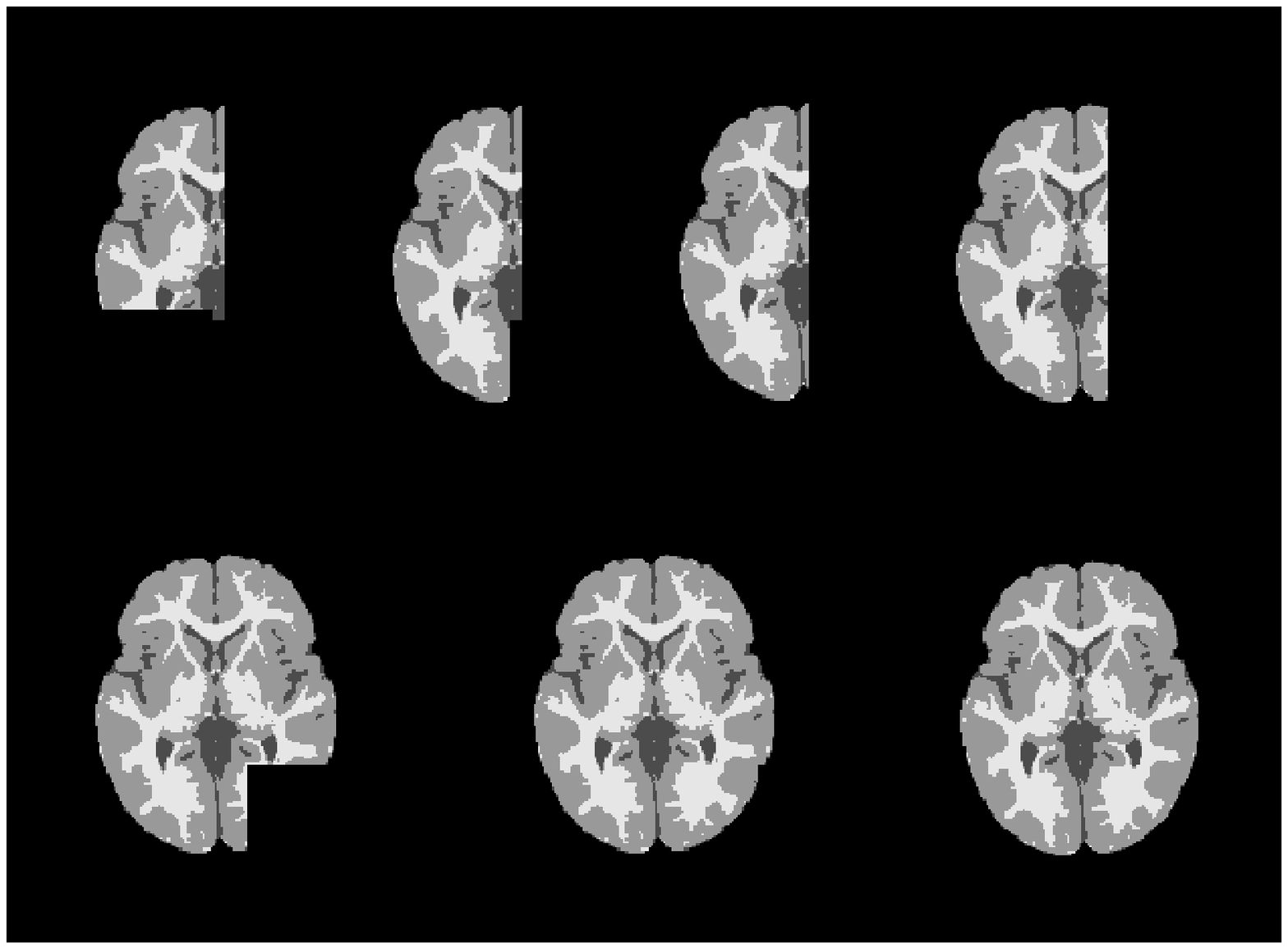}\\
(d)
\caption{(a) Left and right image region overlap, (b) upper and lower image region overlap, (c) Maximum a posteriori estimate of the label configuration $I$ in presence of observations $(I_r,I_l)$, (d) Sequental steps of the stitching algorithm in direction from top left to bottom right.}
\label{st}
\end{figure}

We define an undirected graph $G=(V,E)$ to represent a lattice structure of the image $I_s$, where $V$ are nodes of the graph and $E$ are pairs of neighbouring nodes or edges. For every pixel within $I_s$ there is a hidden node. The node is a random variable taking values from the state space $S=\{1\: \text{(CSF)},\: 2\: \text{(GM)},\: 3\: \text{(WM)}, \:4\: \text{(BG)}\}$. 
The nodes are divided into 2 sets (Figure \ref{mrf}),
\begin{itemize}
\item $Y_{i j}$ - the observed random variable with a label $s \in S$ at a pixel coordinate $(i,j)$,
\item $X_{i j}$ - a hidden node.   
\end{itemize}
Given neighbouring node interactions it is natural to model $Y$ as a realization of a Markov Random Field $X$ with respect to the graph $G=(V,E)$. The hidden nodes $X_{ij}$ and $X_{mn}$ are connected in a lattice way if $i \neq j$ and $m \neq n$. Each possible label coupling and label weights are modelled by the edge potential $\Psi(X_{ij}, X_{mn})$ and the node potential $\Phi(X_{ij})$, respectively. The posterior probability density of the joint label configuration becomes
\begin{equation}
p(X | Y)=\frac{1}{Z(Y)} \prod_{(X_{ij},X_{mn}) \in E} \Psi(X_{ij},X_{mn}) \prod_{(i,j) \in V} \Phi(X_{ij}),
\end{equation}  
where the first product is over all pairs of neighbouring nodes and the second one over all nodes. $Z(Y)$ is a normalizing constant that sums the probabilities over all possible configurations of the variables to 1.
\newline
To find labelling $X$ of the joint region $I_s$ with the highest probability $p(X | Y)$ we estimate $\Psi(X_{ij}, X_{mn})$ and  $\Phi(X_{ij})$ for every node $X_{ij}$ from two sets of observations $Y$ (e.g., $Y=\{I_l, I_r\}$). $\Psi(X_{ij}, X_{mn})$ is defined on $S \times S$ and for every $X_{ij}$ it is given in the form of a $4 \times 4$ matrix with entries
\[
\Psi(X_{i j}=s_k,X_{m n}=s_l)=
\begin{cases}
1,            &\text{if $(Y_{ij}, Y_{mn})_{I_r}=(Y_{ij},Y_{mn})_{I_r}=(s_k, s_l)$};\\
0.5,   &\text{if $(Y_{ij},Y_{mn})_{I_l}=(s_k,s_l)$}\\
& \&\; (Y_{ij},Y_{mn})_{I_r} \neq (s_k,s_l)\;\text{or vice versa};\\
0.01,            &\text {$(s_k,s_l)$ is not observed in $I_r$ and $I_l$}.
\end{cases}
\]
Here, $(s_k, s_l) \in S \times S$, the space of all possible combinations of label pairs. The node potential assigns weight to every possible value of the node $X_{i j}$ as follows
\[
\Phi(X_{ij}=s_k)=
\begin{cases}
w, &\text{if ${Y_{ij}}_{I_l}={Y_{ij}}_{I_r}=s_k$};\\
w \cdot 0.5, &\text{if ${Y_{ij}}_{I_l}=s_k$};\\
&\text{$\& \; {Y_{ij}}_{I_r} \neq s_k$ or vice versa};\\
w \cdot 0.01, &\text{if $s_k$ is not observed in $I_r$ and $I_r$}.
\end{cases}
\]
Here, we also take into account nodal locations, namely, the rightmost nodes in $I_r$ and the leftmost nodes in $I_l$ (similarly, the uppermost nodes in $I_{up}$ and the lowermost nodes in $I_{low}$) are more likely to have values observed in $I_r$ and $I_l$ (or $I_{up}$ and $I_{low}$), respectively. In this way, we ensure the continuity of label propagation from non-overlapping regions to the joint region $I_s$. The additional weight function $w$ is defined as
\[
w(X_{i j})=
\begin{cases}
\frac{3}{4}, &\text{if $X_{i j}$ is the leftmost/rightmost node in $I_l$ /$I_r$}\\
 &\text{(or the uppermost/lowermost node in $I_{up}$/ $I_{low}$ )};\\
\frac{1}{4},& \text{otherwise }
\end{cases}
\] 
With this setup we implemented the iterative Simulated Annealing (SA) algorithm \citep{Grenander1996} to find the most probable joint region for each pair of overlapping classified subimages in a brain slice. We initialized $I_s$ as a region containing the first two columns/rows of $I_l$/$I_{up}$ and the last two columns/rows of $I_r$/$I_{low}$. Such initialization provides a good approximation to the optimal solution and speeds up the convergence of the algorithm.
 \newline
An example of SA application is shown in Figure \ref{st}.c. The leftmost columns of $I_r$ and $I_l$ appear slightly different in presence of the CSF and GM.The rightmost columns of $I_l$ and $I_r$ only differ in the value of a single central pixel. The optimal labelling $I$ preserves the label configuration of its first and last columns as they appear in their respective overlapping regions $I_l$ and $I_r$. 

We started with the upper left subimage and stitched its neigbouring subimages vertically and horizontally. Then for each of the neigbouring subimages we identified overlapping ones in both horizontal and vertical directions, and glued them together using the SA algorithm. By progressive stitching of constituent image parts we assembled the entire brain slice as illustrated in Figure \ref{st}.d. The proposed SA algorithm yields seamless estimates of the joint regions.    
\FloatBarrier
\section{Results and Discussion}
\label{res}
\begin{figure}[h!]
\centering
\includegraphics[width=5.8cm,height=3.5cm]{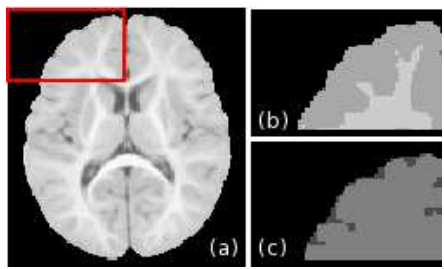}
\caption{(a) T1w reference subdomain (frontal lobe) from the brain template  for ages 8 to 11 months and the corresponding (b) PVE classification and (c) CSF patterns detected by KFDA.}
\label{class1}
\end{figure}

\begin{figure}[h!]
\centering
\includegraphics[width=5.8cm,height=3.5cm]{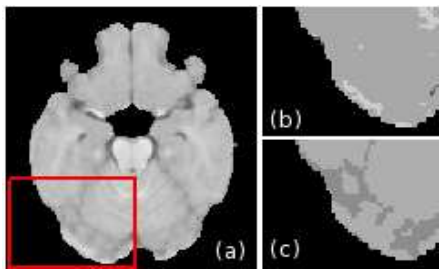}
\caption{(a) T1w reference subdomain (occipital lobe) from the brain template  for ages 8 to 11 months and the corresponding (b) PVE classification and (c) CSF patterns detected by KFDA.
}
\label{class2}
\end{figure}

We applied the local KFDA-based method to segment the brain template for ages 8 to 11 months initially classified into GM, WM and the CSF using PVE. The proposed approach leads to a significant improvement in the CSF detection throughout the brain almost doubling the initial CSF volume (see Figures \ref{class1} and \ref{class2}). The initial CSF cluster determined by PVE consisted of 26717 voxels and has increased to 53681 voxels with application of KFDA. 
\begin{table}[h!]
\begin{center}
\caption{Local comparison of the PVE and KFDA performance via MSSIM.
Pediatric brain template for ages 8 to 11 months. }
\vspace{0.2cm}
\begin{tabular}{| l | c | c || c | c | r |}
\hline
Domain & PVE & KFDA & Domain & PVE & KFDA\\
\cline{1-3} \cline{4-6}
1 & 0.9286 & 0.9206 & 21 & 0.8481 & \cellcolor[gray]{0.7} 0.8848 \\
\cline{1-3} \cline{4-6}
2 & 0.8963 &\cellcolor[gray]{0.7} 0.9000 & 22 & 0.9496 & \cellcolor[gray]{0.7} 0.9595 \\
\cline{1-3} \cline{4-6}
3 & 0.8357 & \cellcolor[gray]{0.7} 0.8500 & 23 & 0.8890 & \cellcolor[gray]{0.7} 0.8975\\
\hline
4 & 0.8682 & \cellcolor[gray]{0.7} 0.8768 & 24 &  0.9226 & 0.9197\\
\hline
5 & 0.9014 & 0.8958 & 25 & 0.8769 & \cellcolor[gray]{0.7}0.8776 \\
\hline
6 & 0.8776 & 0.8709 & 26 & 0.9155 & \cellcolor[gray]{0.7}0.9496 \\
\hline
7 & 0.8623 & \cellcolor[gray]{0.7} 0.8676 & 27 & 0.9502 & \cellcolor[gray]{0.7} 0.9628 \\
\hline
8 & 0.8850 & 0.8753 & 28 & 0.8642 & \cellcolor[gray]{0.7} 0.8675\\
\hline
9 & 0.8337 & \cellcolor[gray]{0.7} 0.8530 & 29 & 0.8978 & 0.8897\\
\hline
10 & 0.8874 & \cellcolor[gray]{0.7} 0.9254 & 30 & 0.8795 & 0.8765 \\
\hline
11 & 0.8375 & 0.8349 & 31 & 0.8261 & \cellcolor[gray]{0.7} 0.8525\\
\hline
12 & 0.8129 & \cellcolor[gray]{0.7} 0.8545 & 32 & 0.8733 & 0.8709 \\
\hline
13 & 0.9027 & \cellcolor[gray]{0.7} 0.9034 & 33 & 0.8722 & \cellcolor[gray]{0.7} 0.8746 \\
\hline
14 & 0.8928 & \cellcolor[gray]{0.7} 0.8978 & 34 & 0.9743 & \cellcolor[gray]{0.7} 0.9759\\
\hline
15 & 0.9399 & 0.9382 & 35 & 0.9135 & \cellcolor[gray]{0.7} 0.9254\\
\hline
16 & 0.9033 & \cellcolor[gray]{0.7} 0.983 & 36 & 0.8501 & \cellcolor[gray]{0.7} 0.8581\\
\hline
17 & 0.9433 & 0.9416 & 37 & 0.8952 & 0.8927\\
\hline
18 & 0.8768 & \cellcolor[gray]{0.7} 0.8828 & 38 & 0.8191 & \cellcolor[gray]{0.7} 0.8379\\
\hline
19 & 0.8828 & \cellcolor[gray]{0.7} 0.8878 & 39 & 0.8198 & \cellcolor[gray]{0.7} 0.8456\\
\hline
20 & 0.8688 & \cellcolor[gray]{0.7} 0.8807 & 40 & 0.9566 & \cellcolor[gray]{0.7} 0.9652\\
\hline 
\multicolumn{1}{ c}{}                                  &
\multicolumn{1}{c }{}                                  &
\multicolumn{1}{c }{}                                   &
\multicolumn{1}{c }{}                                   &
\multicolumn{2}{| c |}{Total MSSIM} \\ 
\cline{5-6}
\multicolumn{1}{ c}{}                                  &
\multicolumn{1}{c }{}                                  &
\multicolumn{1}{c }{}                                   &
\multicolumn{1}{c }{}                                   &
\multicolumn{1}{| c | }{0.8668}                  &
 \multicolumn{1}{ c | }{ \cellcolor[gray]{0.7} 0.87588} \\
\cline{5-6}
\end{tabular}
\end{center}
\label{tb1}
\end{table}
The local MSSIM assessment of the performance of KFDA and PVE given in Table \ref{tb1} shows that KFDA outperforms PVE in 28 subdomains out of 40 and overall. KFDA "loses" to PVE in lower contrast subdomains 5, 15 and 17 and 9 other subdomains within the middle CNR range presumably due to overestimation of the CSF in these subvolumes.  

\begin{figure}[h!]
\centering
\includegraphics[width=9cm,height=3.5cm]{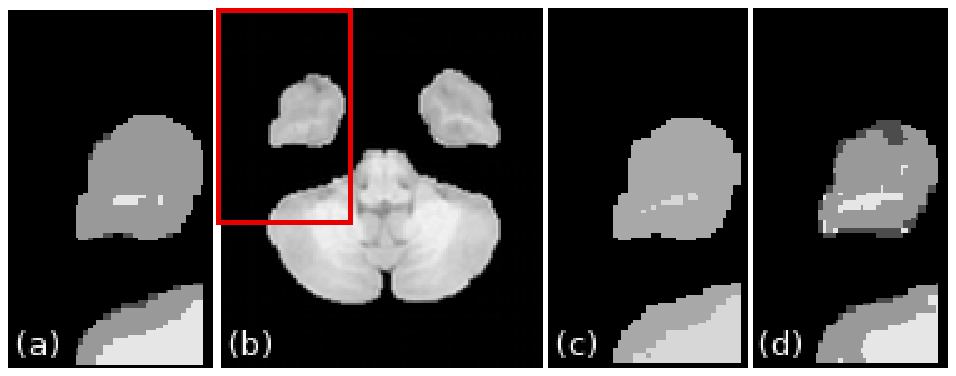}\\

\includegraphics[width=9cm,height=3.5cm]{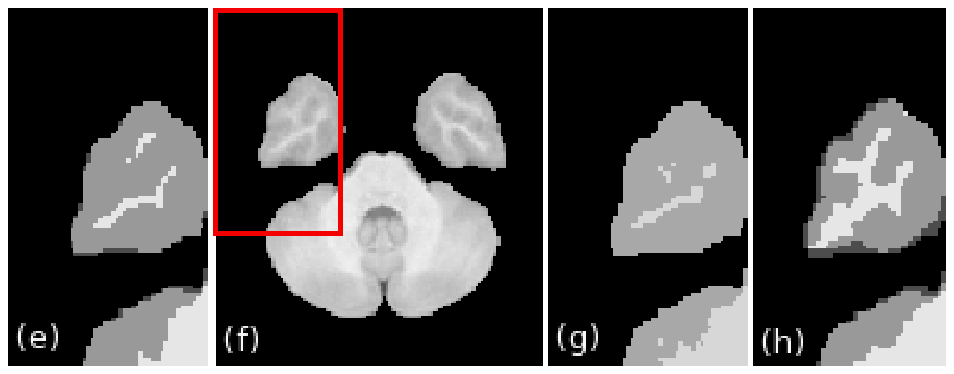}\\

\caption{(a),(e) Initial CSF, GM and WM labels (MSSIM=0.8454) in (b),(f) the T1w reference subdomain (left temporal lobe near the base of the brain) from the brain template  for ages 44 to 60 months and the corresponding (c),(g) PVE (MSSIM=0.8419) and (d),(h) KFDA (MSSIM=0.8911) classifications. MSSIM was computed over 14 slices of the brain subdomain enclosed in the red frame.}
\label{sl1}
\end{figure}

We also applied our segmentation method to the 3D brain template for ages 44 to 60 months with initialization obtained by a label transfer from the brain template for ages 4.5 to 8.5 years.  Figure \ref{sl1} illustrates the performance of KFDA using classified brain slices. 
\begin{table}[h]
\begin{center}
\caption{Local comparison of the initial label transfer, PVE and KFDA classifications via MSSIM. 
Pediatric brain template for ages 44 to 60 months.}
\vspace{0.2cm}
\begin{tabular}{| l | c | c | c || c | c | c | r |}
\hline
Domain & Initial & PVE & KFDA & Domain & Initial & PVE & KFDA\\
\hline
1 & 0.7809 & 0.7846 & \cellcolor[gray]{0.7} 0.8243 & 12 & 0.8444 & 0.8856 & 0.8718\\
\hline
2 & 0.8051 & 0.8454 & 0.8389 & 13 & 0.7904 & 0.8247 & \cellcolor[gray]{0.7} 0.8320\\
\hline
3 & 0.8405 & 0.8405 & \cellcolor[gray]{0.7} 0.8759 & 14 & 0.8142 & 0.8304 & \cellcolor[gray]{0.7} 0.868\\
\hline
4 & 0.7815 & 0.8212 & \cellcolor[gray]{0.7} 0.8422 & 15 & 0.8560 & 0.8804 & \cellcolor[gray]{0.7} 0.8825\\
\hline
5 & 0.8230 & 0.8671 & 0.8601 & 16 & 0.8004 & 0.8231 & \cellcolor[gray]{0.7} 0.8467\\
\hline
6 & 0.7422 & 0.7422 & \cellcolor[gray]{0.7} 0.8146 & 17 & 0.7785 & 0.7785 & \cellcolor[gray]{0.7} 0.8292\\
\hline
7 & 0.7071 & 0.7384 & \cellcolor[gray]{0.7} 0.7830 & 18 & 0.7755 & 0.8234 & \cellcolor[gray]{0.7} 0.8479\\
\hline
8 & 0.8266 & 0.8864 & 0.8676 & 19 &  0.7600 & 0.7600 & \cellcolor[gray]{0.7} 0.8373\\
\hline
9 & 0.9257 & 0.9482 & 0.9400 & 20 & 0.7213 & 0.7299 & \cellcolor[gray]{0.7} 0.8058\\
\hline
10 & 0.7925 & 0.8511 & 0.8348 & 21 & 0.8347 & 0.8571 & \cellcolor[gray]{0.7} 0.8747\\
\hline
11 & 0.7529 & 0.8373 & 0.8128 & 22 & 0.7416 & 0.8071 & \cellcolor[gray]{0.7}  0.8251\\
\hline
\multicolumn{1}{ c}{}                                  &
\multicolumn{1}{c }{}                                  &
\multicolumn{1}{c }{}                                   &
\multicolumn{1}{c }{}                                   &
\multicolumn{1}{| c |}{means} &
\multicolumn{1}{c |}{0.7952} &
\multicolumn{1}{c |}{0.8256} &
\multicolumn{1}{ c |}{0.8458} \\
\cline{5-8}
\end{tabular}
\end{center}
\label{tb44}
\end{table}
Local MSSIM assessment of the quality of classifications by initial label transfer, KFDA and PVE classifications suggests that KFDA improves initial classification and outperforms PVE in all but 7 brain subvolumes (Table \ref{tb44}). However, in high contrast subvolumes KFDA tends to overestimate WM and CSF leading to poorer structural similarity estimates. Overall, in comparison to PVE, KFDA detects brain tissue classes throughout the entire brain with a higher accuracy.   
\FloatBarrier
This example demonstrates the capability of the proposed method to identify small brain structures hardly visible in low contrast subdomains of reference images.
\section{Summary}
\label{disc}
In this paper, we developed a powerful KFDA-based framework that overcomes methodological limitations of existing segmentation approaches in infant brain MRI such as global modelling of tissue intensity distributions and dependency on probabilistic brain atlas. The proposed method
\begin{itemize}
\item uses a general class separability criterion since it does not impose tissue intensity distribution models on the input data and captures a wide range of tissue cluster non-linearities,
\item avoids a build-up of various techniques for refinement of segmentation results and accommodation of intra-class intensity variability,
\item takes advantage of multi-channel brain imaging data and avoids inconsistencies that arise when segmenting each image type separately,
\item classifies MR brain data locally and improves detection of spatial tissue patterns in low contrast subdomains,
\item adapts initial classification to the tissue intensity distributions within an individual brain scan,
\item allows comparison of classification algorithms and automatic monitoring of the quality of classification in the absence of the ground truth via the objective image quality metric SSIM.
\end{itemize}
Our framework is semi-supervised since we used tissue label transfer from an older brain MRI to a younger brain anatomy for initialization. The initialization provides the best guess on the spatial locations of tissue classes that is updated in accordance with tissue intensity histograms of brain imaging data. 
\newline
We explored the potential of KFDA in applications to brain tissue classification of infant brain MRI, in particular, the NIH pediatric database. We observed that even with poor initial estimates of the class clusters in the brain template for ages 8 to 11 months KFDA compensates for the underestimates and detects most of the tissues visible in MRI. Overall, application of the KFDA-based method yields a more accurate quantification of brain tissue volumes from infant brain MRI.
 \newline
The proposed method is mathematically rich and offers avenues for the further advancement of KFDA-based methodology. To mention a few,  
\begin{itemize}
\item The classification problem in the absence of ground truth can be stated in a mathematically rigorous way, namely, as an optimization of MSSIM over a set of overlapping voxels in the stereotaxic space. The Gibbs field model of the unknown label configuration defined on this set with the Gibbs interaction energy in the form of the negative of MSSIM increment will allow optimal estimation of labelling via SA. The classification results obtained in this way do not correspond to the global maximum of MSSIM.    
\item A more competitive perception quality measure can be used for the evaluation of the classification quality. The information content-weighted SSIM \citep{Wang2011} is an advanced version of SSIM that assigns local information content weights to image components containing more information. Such a metric will emphasize tissue boundaries in the classified image and measure the structural similarity between classified and T1w(reference) images more accurately.
\item A more general kernel function in the form of a linear combination of Gaussians with different bandwidths can be explored.  
\item Further understanding of variation in tissue class estimates between neighbouring subimages resulting from the choice of the width of the overlapping area may lead to seamless brain assembly \citep{Pelletier2014}, \citep{Koen2010} thus avoiding the need for SA application.     
\end{itemize}
The proposed method is applicable to brain tissue classification of multichannel brain MRI for ages 8 months and older. 

\section{Acknowledgements}
This project was funded in whole or in part by the Montreal Neurological Institute in the form of a postdoctoral fellowship, the National Institute of Child Health and Human Development, the National Institute on Drug Abuse, the National Institute of Mental Health, and the National Institute of Neurological Disorders and Stroke (Contract \#s N01-HD02-3343, N01-MH9-0002, and N01-NS-9-2314, -2315, -2316, -2317, -2319 and -2320). This manuscript expresses the views of the authors and may not reflect the opinions or views of the NIH. 

 \appendix
 \section{Fundamentals of KFDA}
 \label{A1}
Originally, Kernel Fisher Discriminant Analysis was proposed for a binary classification problem \citep{Mika1999} and later it has been extended to multiclass classification. Presented below is a classical binary KFDA.
\newline
Let $X$ be an input set, an arbitrary subset in $R^n$ and let $Y=\{-1,+1\}$ be class label set. We refer to an input pair $(x,y)$, where $x \in X$ and $y \in Y$ as a sample. The assumption is that all samples are drawn randomly and independently from unknown probability distribution over $X \times Y$.
Let $X_{-}=\{x_1,x_2,...,x_{l_1}\}_{-}$ and $X_{+}=\{x_{l_1+1},x_{l_1+2},...,x_{l_1+l_2}\}_{+}$ be samples from two different classes (referred to as negative and positive for a reason that will become clear later) with means $\mu_{-}$ and $\mu_{+}$.  
\newline
KFDA finds a non-linear direction of maximal information discrimination in the input space by first mapping the data $X_{-}$ and $X_{+}$ non-linearly into a high-dimensional  \textit{feature space} $\mathcal{H}$ through an implicit mapping $\phi: X \rightarrow \mathcal{H}$ and computing an optimal separating hyperplane there that corresponds to a non-linear separating surface in the input space. The power of KFDA lies in ability to capture complex non-linear structures of clusters in the input space.
\newline
To understand a kernel "trick" associated with KFDA we first formulate a linear discriminant optimality criterion. In the context of our classification problem in 3D input space, we aim to find the direction $\vec{w}$ of input data projection $\vec{w}^T \cdot \vec{x} \in R^1$ such that    
 \begin{itemize}
 \item between-class variance $\vec{w}^T S_{B} \vec{w}$ is maximized, where 
 \begin{equation*}
 S_{B}=(\mu_{-}-\mu_{+})(\mu_{-}-\mu_{+})^T,
 \end{equation*} 
 \item within-class variance $\vec{w}^T S_{W} \vec{w}$ is minimized,  where 
 \begin{equation*}
 S_W=\sum_{i \in \{-1,+1 \}}\sum_{x \in X_i}(x-\mu_i)(x-\mu_i)^T.
 \end{equation*}
  \end{itemize}    
 Overall, the objective is to maximize a linear discriminant 
  \begin{equation*}
  J(w)=\frac{\vec{w}^T S_{B} \vec{w}}{\vec{w}^T S_{W} \vec{w}}.
  \end{equation*}   
To find an optimal non-linear direction of data projection, we first non-linearly transform the data with the implicit mapping $\phi$ to the abstract feature space $\mathcal{H}$ and compute the linear discriminant in this feature space. Then, in $\mathcal{H}$ we have
\begin{align}
\mu_{-}^{\phi} &=\frac{1}{l_1}\sum_{i=1}^{l_1} \phi({x_i}_{-}),\:\:\:\mu_{+}^{\phi}=\frac{1}{l_2}\sum_{i=1}^{l_2} \phi({x_i}_{+}),
\label{mu}
\\ 
S_{B}^{\phi} &=(\mu_{-}^{\phi}-\mu_{+}^{\phi})(\mu_{-}^{\phi}-\mu_{+}^{\phi})^T,
\label{sb}
\\
S_W^{\phi} &=\sum_{i \in \{-1,+1 \}}\sum_{x \in X_i}(\phi(x)-\mu_i^{\phi})(\phi(x)-\mu_i^{\phi})^T.
\label{sw}
\end{align}
Thus, the linear discriminant in $\mathcal{H}$ is
\begin{equation}
  J(w)=\frac{\vec{w}^T S_{B}^{\phi} \vec{w}}{\vec{w}^T S_{W}^{\phi} \vec{w}}.
\label{fisher}
\end{equation}
Since $\mathcal{H}$ is a high-dimensional space whose dimension would be equal to the total number of interior brain voxels in applications to MR brain images it is impossible to solve \ref{fisher} directly. We seek a formulation of the optimality criterion in terms of dot-products $<\phi(x),\phi(z)>_H$ of the mapped samples $x,z \in X$ since by Mercer's theorem \citep{ShaweTaylor2004} we can compute these dot-products without explicit knowledge of the mapping $\phi$ via kernel functions. The kernel $K: X \times X \rightarrow R$ is a symmetric function that defines the dot product $K(x,z)={<\phi(x),\phi(z)>}_H$ for $\forall x, z \in X$. Possible choices for $K$ are Gaussian radial basis function(RBF), $K(x,z)=\exp(-\frac{\|\phi(x)-\phi(z)\|^2}{c})$, polynomial kernels $k(x,z)=(<\phi(x),\phi(z)>)^d$ and sigmoidal functions $k(x,z)=\tanh(a <\phi(x),\phi(z)>+b)$.  
\newline
Therefore, we make use of the kernel trick and rewrite (\ref{fisher}) in computable form. Let the total number of samples be $l=l_1+l_2$.  Given that any solution $ w \in H$ lies in the span of training samples $\{\phi(x_i)\}, i=1,2,...,l$ \citep{Saitoh1988}
\begin{equation}
w=\sum_{i=1}^l\alpha_i \phi(x_i).
\label{exp}
\end{equation}   
Using equations (\ref{exp}) and (\ref{mu}) we obtain
\begin{equation}
w^T \mu_{-}^{\phi}=\frac{1}{l_1}\sum_{j=1}^l\sum_{k=1}^{l_1}\alpha_j K(x_j,{x_k}_{-})=\alpha^T M_{-},
\label{wmu}
\end{equation}
where $M_{-}$ is an $l \times1$ vector with entries ${M_{-}}_j=\frac{1}{l_1}\sum_{k=1}^{l_1}K(x_j,{x_k}_{-})$.
Similarly, we find
\begin{equation}
w^T\mu_{+}^{\phi}=\alpha^T M_{+}.
\label{wmu2}
\end{equation}
By using equations (\ref{sb}), (\ref{wmu}) and (\ref{wmu2}) the numerator of $J(w)$ in (\ref{fisher}) can be rewritten as
\begin{equation}
w^T {S_B} ^{\phi}w=\alpha^T (M_{-}-M_{+})(M_{-}-M_{+})^T \alpha=\alpha^T M \alpha.
\label{sb1}
\end{equation} 
By using equations (\ref{mu}), (\ref{exp}) and (\ref{sw}) the denominator of $J(w)$ in (ref{fisher}) becomes
\begin{equation}
w^T{S_W}^{\phi}w=\alpha^T [k_{-}(I-\textbf{1}_{l_1})k_{-}^T+k_{+}(I-\textbf{1}_{l_2})k_{+}^T] \alpha=\alpha^T N \alpha,
\label{sw1}
\end{equation} 
where $k_{-}$ and $k_{+}$ are kernel matrices of sizes $l \times l_1$ and $l \times l_2$, respectively, with entries $(k_m)_{i j}=K(x_i,{x_j}_m), m \in \{-1,+1\}$, $I$ is the identity matrix of size $l_n \times l_n$ and $\textbf{1}_{l_n}$ is the matrix with all entries $\frac{1}{l_n}, n=1,2$.  
 Combining (\ref{sb1}) and (\ref{sw1}) we aim to find a vector $\alpha$ in $\mathcal{H}$ that maximizes
 \begin{equation}
 J(\alpha)=\frac{\alpha^T M \alpha}{\alpha^T (N +\beta I)\alpha}. 
 \label{fin}
 \end{equation} 
 Here, $\beta$ is a small positive regularization parameter added to the variance entries of the within-class variance-covariance matrix $N$ to ensure its positive-definiteness since N may be singular. The decision boundary between the classes in the input space is a set of points that satisfies
 \begin{equation}
 h(x)=\sum_{i=1}^{l} \alpha_i K(x_i,x)+b=0,
\text{ where $b$ is an offset.}
\end{equation} 
Since $h(x)$ depends on the kernel choice, the classification result will also be kernel-dependent. As such, it is important to choose the kernel function that best captures a non-linear behaviour of class clusters based on their natural occurrences in the input space.

\bibliographystyle{apalike}
\bibliography{mybiblio}
\nocite{*}


\end{document}